\documentclass[fleqn,usenatbib]{mnras}

\usepackage{newtxtext,newtxmath}

\usepackage[T1]{fontenc}
\usepackage{color,soul}

\usepackage{tikz}
\usetikzlibrary{shapes.geometric, arrows}

\DeclareRobustCommand{\VAN}[3]{#2}
\let\VANthebibliography\thebibliography
\def\thebibliography{\DeclareRobustCommand{\VAN}[3]{##3}\VANthebibliography}

\newcommand{\beorn}{{\tt BEoRN\,}}

\usepackage{graphicx}	
\usepackage{amsmath}	

\title[BEoRN]{{\tt BEoRN}: A fast and flexible framework to simulate the epoch of reionisation and cosmic dawn}
\author[T. Schaeffer et al.]{
Timoth\'ee Schaeffer$^{1}$\thanks{E-mail: timothee.schaeffer@uzh.ch}, 
Sambit K. Giri$^{1,2}$,  and 
Aurel Schneider$^{1}$
\\
$^{1}$Center for Theoretical Astrophysics and Cosmology, Institute for Computational Science, University of Zurich,\\
Winterthurerstrasse 190, CH-8057 Z{\"u}rich, Switzerland\\
$^{2}$Nordita, KTH Royal Institute of Technology and Stockholm University, Hannes Alf\'vens v\"ag 12, SE-106 91 Stockholm, Sweden
}

\date{Accepted XXX. Received YYY; in original form ZZZ; Report Number NORDITA 2023-019}

\pubyear{2023}

\begin{document}
\label{firstpage}
\pagerange{\pageref{firstpage}--\pageref{lastpage}}
\maketitle

\begin{abstract}
In this study, we introduce {\tt BEoRN} (Bubbles during the Epoch of Reionisation Numerical Simulator), a publicly available Python code that generates three-dimensional maps of the 21-cm signal from the cosmic dawn and the epoch of reionisation. Built upon $N$-body simulation outputs, {\tt BEoRN} populates haloes with stars and galaxies based on a flexible source model. It then computes the evolution of Lyman-$\alpha$ coupling, temperature, and ionisation profiles as a function of source properties, and paints these profiles around each source onto a three-dimensional grid. The code consistently deals with the overlap of ionised bubbles by redistributing photons around the bubble boundaries, thereby ensuring photon conservation. It accounts for the redshifting of photons and the source look-back effect for the temperature and Lyman-$\alpha$ coupling profiles which extend far into the intergalactic medium to scales of order 100 cMpc. We provide a detailed description of the code and compare it to results from the literature. After validation, we run three different benchmark models based on a cosmological $N$-body simulation. All three models agree with current observations from UV luminosity functions and estimates of the mean ionisation fraction. Due to different assumptions regarding the small-mass stellar-to-halo relation, the X-ray flux emission, and the ionising photon escape fraction, the models produce unique signatures ranging from a cold reionisation with deep absorption trough to an emission-dominated 21-cm signal, broadly encompassing the current uncertainties at cosmic dawn.
The code {\tt BEoRN} is publicly available at  \url{https://github.com/cosmic-reionization/BEoRN}.

\end{abstract}




\section{Introduction}\label{intro}

The hyperfine transition of neutral hydrogen generates photons at the wavelength of 21 cm, opening a new observational window into the early Universe approximately one billion years after the Big Bang. During this era, the radiation from the first stars and galaxies pushes the spin temperature out of equilibrium before heating and eventually ionising the neutral hydrogen of the intergalactic medium (IGM). Next to the source properties, the 21-cm signal depends on the clustering and temperature distribution of the neutral gas, the primordial background radio emission, and the detailed interaction processes between radiation and matter. It is therefore not surprising that the 21-cm radiation from the cosmic dawn contains a wealth of information about the properties of the first stars \citep{Fialkov_2014,Mirocha:2017xxz,Ventura:2022rwn,Sartorio2023}, galaxies \citep{Park_18,Reis_2020,Hutter_2021}, and black holes \citep{Pritchard_2007,Ross_2019_QSO_contribution}. It can furthermore be used to constrain the cosmological model \citep{Liu_2016,HaloModel_reio,Shmueli} and, in particular, the dark sector, such as the nature of dark matter \citep{Sitwell_2014,Chatterjee2019,Nebrin2019,Munoz_2020_probing_small_scales_21cm,Jones2021,Giri_22,Hotinli_22,Flitter_22,Hibbard_2022}, interactions between the dark and visible sector \citep{Barkana_2018,Fialkov_2018,Kovetz_2018,Lopez-Honorez:2018ipk,Mosbech:2022uud}, or potential exotic decay and annihilation processes \citep{D'Amico_2018,Liu_18,Mitridate_189}.

Reliable detection of the 21-cm signal at these redshifts has yet to be achieved, but ongoing experiments, such as the 
Giant Metrewave Radio Telescope \citep[GMRT,][]{GMRT_2013_up_lim}, the Precision Array for Probing
the Epoch of Reionization \citep[PAPER,][]{PAPER_2019_uplim}, the Murchison Widefield Array \citep[MWA,][]{MWA_2020_z6.5_8.7}, the Low-Frequency ARray \citep[LOFAR,][]{LOFAR_z9}, and the Hydrogen Epoch of Reionization Array \citep[HERA,][]{HERA_22b} have provided upper limits on the 21-cm power spectrum for a broad range of redshifts. These bounds have been used to exclude regions of the parameter space describing extreme properties of the IGM during the epoch of reionisation \citep{LOFAR_constraints_Ghara_2020,MWA_constraints_Ghara_21,MWA_constr_Greig,Greig_21_LOFAR_interpreting,HERA_interpretation_Abdurashidova_2022}.


The Square Kilometre Array (SKA), a next-generation radio interferometer, is currently under construction in South Africa and Western Australia. Its low-frequency component, SKA-low, has the capability to not only measure the 21-cm power spectrum with high signal-to-noise ratio but also provide sky images at redshifts around $z\approx 5-25$ \citep[e.g.][]{mellema2015hi,2015_Wyithe_Imaging,ghara2017imaging,giri2018bubble,bianco2021deep}. The potential of SKA-low for studying the cosmic dawn and reionization era has been extensively investigated in various studies, exploring properties of the ionizing sources and the ionization structure of the universe \citep[e.g.][]{giri2018optimal,zackrisson2020bubble,giri2021measuring,gazagnes2021inferring,bianco2023deep}. These studies highlight the significant role that SKA-low will play in advancing our understanding of these critical cosmic epochs.

Next to the tremendous experimental effort, accurate and reliable theoretical methods to model the 21-cm signal at the required accuracy level are currently being developed.
Modelling the 21-cm signal is challenging as it involves a broad dynamical range from minihaloes to cosmological scales. It depends on the details of hydrodynamical feedback processes for galaxies, the propagation of radiation through large cosmological scales, and the detailed interaction processes of photons with gas particles of the IGM \citep[e.g.,][]{iliev2006simulating,mellema2006simulating,trac2007radiative}. 

One option is to predict the 21-cm signal with the help of coupled radiative-transfer hydrodynamic simulations, some well-known examples being the Cosmic Dawn (CoDA) \citep{CoDA_1,ocvirk2020cosmic,CoDA_2}, the 21SSD \citep{Semelin_LICORICE}, and the THESAN simulations \citep{Thesan_1,Thesan_2}. Another option is to post-process N-body simulations with ray-tracing algorithms, such as the Conservative, Causal Ray-tracing code \citep[{\tt C$^2$RAY};][]{MELLEMA_2006} or the Cosmological Radiative transfer Scheme for Hydrodynamics \citep[{\tt CRASH};][]{maselli2003crash}. Full radiative-transfer numerical methods are fundamental to understanding the 21-cm signal and estimating the accuracy of more approximate methods. However, they are very computationally expensive and can hardly be used to scan the vast cosmological and astrophysical parameter space. To perform Bayesian inference analysis on a mock 21-cm data set, semi-numerical algorithms are often used, better suited to generate thousands of realizations of the signal itself. They rely on the excursion set formalism \citep{Exc_Set_Furla_2004_TheGrowthHIIRegions}, such as {\tt 21cmFAST} \citep{21cmFAST_2010} or {\tt SIMFAST21} \citep{SimFast21}.


In this paper, we present the new framework {\tt BEoRN} which stands for \emph{Bubbles during the Epoch of Reionisation Numerical simulator}. The code is based on a one-dimensional radiative transfer method in which interactions between matter and radiation are treated in a spherically symmetric way around sources. This approach is significantly faster than full 3-d radiative transfer codes and arguably more precise than semi-numerical algorithms which are not based on individual sources. In this aspect, {\tt BEoRN} is similar to other existing codes such as {\tt BEARS} \citep{thomas2009fast} or {\tt GRIZZLY} \citep{Grizzly_comparison_3D_1D}. However, in contrast to other 1d radiative transfer codes, {\tt BEoRN} self-consistently accounts for the evolution  of individual sources during the emission of photons. This includes both the redshifting of photons due to the expansion of space and the increase of luminosity caused by the growth of individual sources over time. Both effects have a non-negligible influence on the radiation profile surrounding sources.

The {\tt BEoRN} framework allows for a flexible parametrisation to model any source of radiation, such as e.g. Pop-III stars, galaxies, or quasars. It produces a 3-dimensional (3D) light-cone realisation of the 21-cm signal from the cosmic dawn to the end of reionisation including redshift space distortion effects. The underlying gas density field as well as the position of sources is directly obtained from outputs of an $N$-body simulation. We have designed {\tt BEoRN} to be user-friendly and modular so that it can be applied in combination with different gravity solvers or source models, for example.


The paper is structured as follows: Section~\ref{sec2} describes the {\tt BEoRN} code, while section~\ref{sec3} validates it by comparing its predictions with the publicly available {\tt 21cmFAST} code. In section~\ref{sec:results}, three benchmark models are presented, calibrated to the latest observations, and the evolution of the 21-cm signal during the cosmic dawn and epoch of reionization is studied. The work concludes with a summary and conclusion in section~\ref{sec:conclusion}. 

Note that throughout the paper, physical distance units are specified with the prefix "$p$", while co-moving distance units are specified with the prefix "$c$".
The cosmological parameters used in this work are consistent with Planck 2018 results \citep{Planck_2018}, namely matter abundance $\Omega_{\rm m}=0.31$, baryon abundance $\Omega_{\rm b}=0.045$, and dimensionless Hubble constant $h=0.68$. The standard deviation of matter perturbations at 8$h^{-1}$ cMpc scale is $\sigma_{\rm 8}=0.81$.

\section{Modelling the 21-cm signal}\label{sec2}

\subsection{21-cm differential brightness temperature}\label{sec2:1}

Radio interferometers measure the 21-cm signal against the Cosmic Microwave Background (CMB) radiation. This so-called differential brightness temperature ($dT_{b}$) is a function of position ($\mathbf{x}$) and redshift ($z$) given by \citep[e.g.][]{Cosmo_low_freq_FURLA2006}
\begin{equation}
\begin{aligned}
    dT_{b}(\mathbf{x},z) = {} & 27(1-x_{\rm HII}(\mathbf{x},z))(1+\delta_{\rm b}(\mathbf{x},z))\left(\frac{0.15}{\Omega_{m}h^{2}}\frac{(1+z)}{10}\right)^{\frac{1}{2}}  \\
            & \times \left(\frac{\Omega_{ b}h^{2}}{0.023} \right) \frac{x_{\rm tot}(\mathbf{x},z)}{1+x_{\rm tot}(\mathbf{x},z)} \left(1-\frac{T_{ \rm \gamma}(z)}{T_{\rm k}(\mathbf{x},z)}\right) ~~[\rm mK],
  \label{eq:dTb formula}
\end{aligned}
\end{equation}
where $T_{\gamma}$ is the cosmic microwave background (CMB) temperature, $T_{\rm k}$ is the kinetic temperature of the gas, $x_{\rm HII}$ is the local fraction of ionised hydrogen, and $\delta_{\rm b}$ the baryon overdensity. The total coupling coefficient is defined as the sum of the collisional and Lyman-$\alpha$ coupling coefficients, i.e., $x_{\rm tot} = x_{c}+x_{\alpha}$. We compute them using the following expressions:
\begin{equation}
  x_{\rm \alpha}(\mathbf{x},z) = \frac{1.81\times 10^{11}}{(1+z)}S_{\alpha}J_{\alpha}(\mathbf{x},z)
  \label{eq:xal},
\end{equation}
and
\begin{equation}
  x_{\rm c}(\mathbf{x},z) = \frac{T_{*}}{A_{10} T_{\gamma}(z)}\underset{i}{\sum} n_{i}(z)\kappa_{10}^{i}(T_{\rm k}),
  \label{eq:xcol}
\end{equation}
where $S_{\alpha}$ is given by Eq.~(55) in \citet{Cosmo_low_freq_FURLA2006}  and $J_{\alpha}(\mathbf{x},z)$ is the local flux of Lyman$-\alpha$ photons in $[\rm pcm^{-2}sr^{-1}Hz^{-1}s^{-1}]$. 
$\kappa^{i}_{10}$ $[\rm cm^{3}s^{-1}]$ is the rate coefficient for spin de-excitation in collisions with species $i$, and depends on the local temperature of the gas, and hence, on the position (\textbf{x}). The sum is done over $i=\{e^{-},H\}$, i.e. hydrogen atoms and free electrons, with densities $n_{i} [\rm cm^{-3}]$. $A_{10}[\rm s^{-1}]$ is the Einstein coefficient for spontaneous emission, and $T_{*} = 68$ mK the temperature associated with the hyperfine transition.

The peculiar velocity of the gas in the IGM will modulate the observed frequency of the 21-cm signal \citep[see e.g. eq. 6 in][]{ross2021redshift}. This phenomenon, known as redshift space distortion \citep[RSD][]{Kaiser_1987}, will significantly impact the signal, see e.g. \citet{mao2012redshift,jensen2013probing} and \citet{ross2021redshift} for a detailed study on this. The observed data by radio interferometers will contain the 21-cm signal spatially distributed in the sky and evolving along with observed frequency or redshift. This 3D data is known as the lightcone.
We account for RSD using the scheme described in \citet{jensen2013probing} and create the 21-cm signal lightcone using the method in \citet{datta2012light}, which are implemented in the publicly available package {\sc Tools21cm} \citep{giri2020tools21cm}.


\begin{figure}
    \centering
    \includegraphics[width =1\textwidth,trim=2.8cm 9.cm 0cm 5.0cm,clip]{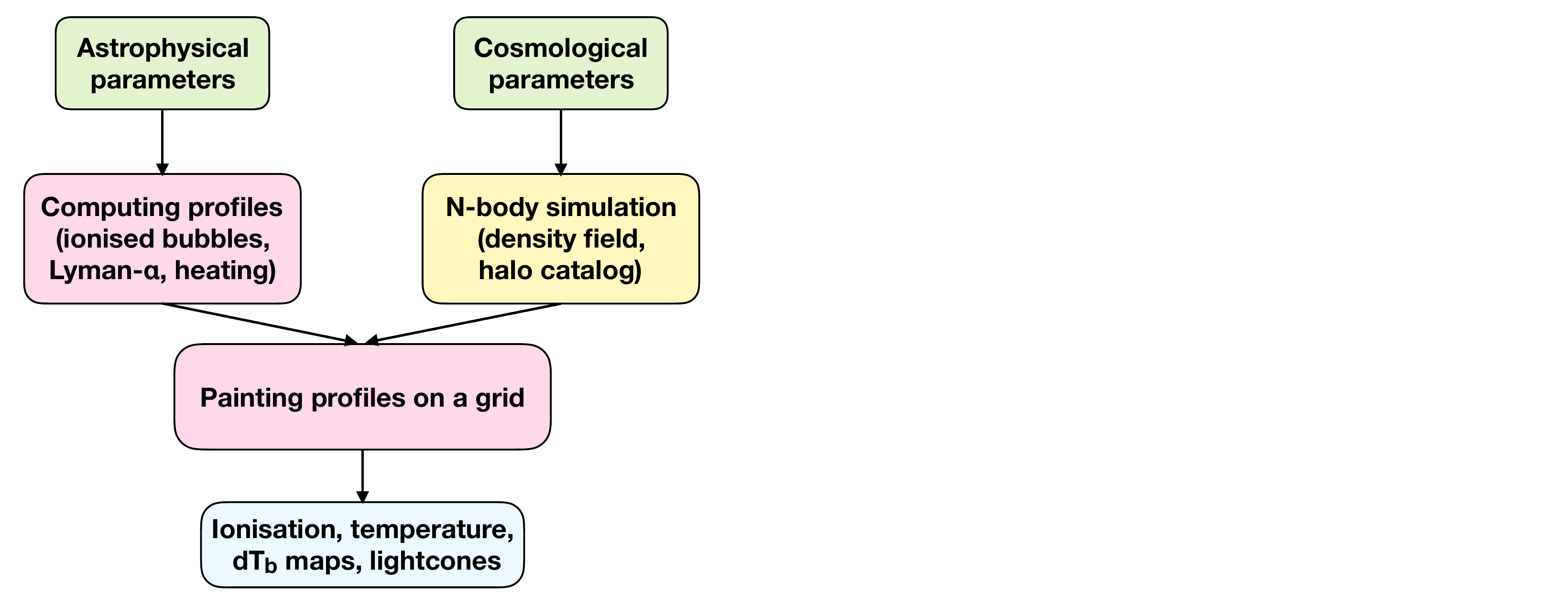}
    \caption{General structure of the {\tt BEoRN} code. On the one hand, the astrophysical source parameters determine the size of ionised bubbles, the gas temperature profiles and the Lyman-$\alpha$ flux profiles. On the other hand, the cosmological parameters control the properties of the matter density field and the abundance of DM haloes, obtained from N-body simulations. The profiles are painted on a 3D grid, centred on DM haloes. The end-product of the code are ionisation, kinetic temperature and 21-cm brightness temperature maps including redshift space distortions, as well as lightcones. 
    }
    \label{fig:Flowchart}
\end{figure}

\subsection{Methodology of {\tt BEoRN}}\label{sec2:2}
The objective of {\tt BEoRN} is to produce brightness temperature ($dT_{b}$) maps on 3D grids, from the beginning of cosmic dawn to the end of the reionisation process. We thereby assume spherical symmetry around sources regarding the photon emission, the gas heating and the Lyman-$\alpha$ coupling processes. In this way we can describe the propagation of photons and the radiation-matter interactions with a set of one-dimensional profiles, greatly simplifying our analysis. Different profiles may overlap leading to complex, non-symmetrical patterns around sources. 

While the temperature and Lyman-$\alpha$ coupling are additive in the sense that profiles from different sources acting on one gas volume can simply be summed up, this is not the case for the ionisation process. In this case, overlapping profiles have to be corrected by redistributing excess photons in order to guarantee photon number conservation. Note that such an approach has been shown to provide accurate patterns of ionisation \citep{Grizzly_comparison_3D_1D}.

We summarise the basic algorithm of {\tt BEoRN} in the following steps:
\begin{enumerate}
\item Assuming an astrophysical source model, {\tt BEoRN} first computes the evolution of 1D ionisation, Lyman-$\alpha$ and temperature profiles for a discrete set of halo masses (between $M\simeq 10^5$ and $10^{15}$ M$_{\odot}$). It thereby assumes haloes to grow according to a universal Mass Accretion Rate (MAR) model from the cosmic dawn ($z>30$) to the end of reionisation ($z<6$).
\item As a next step, these pre-computed spherical profiles are painted onto a 3d grid. They are centred around sources which are assumed to inhabit haloes obtained from a pre-run $N$-body simulation. Regarding the gas density field, {\tt BEoRN} does not assume profiles but directly uses the outputs from the $N$-body simulations.
\item In order to account for the overlap of ionised bubbles, the ``overionised'' cells are redistributed to the bubble outskirts, ensuring photon conservation.
\item The resulting $dT_{b}$ signal is obtained by combining the temperature ($T_{k}$), ionisation ($x_{\rm HII}$), and Lyman-$\alpha$ coupling ($x_{\alpha}$) maps for each redshift according to Eq.~(\ref{eq:dTb formula}).
Finally, all outputs are corrected for the RSD effect and combined to obtain a light cone prediction of the 21-cm signal.
\end{enumerate}

In the next sections, we present the different ingredients of the framework in more detail. In particular we discuss the underlying N-body simulations (Sec.~\ref{sec2:3}), the implementation of the source model (Sec.~\ref{sec2:4}), the calculation of the source profiles (Sec.~\ref{sec2:5}), the painting algorithm (Sec.~\ref{sec2:6}), and the redistribution of ionising photons to ensure photon conservation (Sec.~\ref{sec2:7}).

\subsection{Density field and dark matter haloes}\label{sec2:3}
As initial input, the {\tt BEoRN} framework requires a 3D density field and a corresponding halo catalogue (with positions) for different redshifts. We use outputs from $N$-body simulations run with {\tt Pkdgrav3} \citep{Potter2017} but other methods are applicable as well. For example, a significant speed-up of the code could be achieved by using density fields from Lagrangian perturbation theory combined with halo catalogues from excursion set modelling. Another option would be to couple {\tt BEoRN} with fast and approximate gravity codes such as {\tt COLA} \citep[COmoving Lagrangian Acceleration;][]{tassev2015scola} or {\tt FAST-PM} \citep{FAST_PM}.

Our algorithm assumes that sources reside at the center of dark matter haloes and that baryonic gas tracks dark matter such that their overdensities are identical ($\delta_{\rm b}=\delta_{\rm m}$). These assumptions are expected to be valid for the scales of interest, especially for the most relevant wave modes $0.1\lesssim k\lesssim 3$ Mpc$^{-1}$.


\subsection{Source modelling
}\label{sec2:4}

In this section, we present the source parameterisation used in {\tt BEoRN} which is motivated by the model presented in \citet{HaloModel_Paper,HaloModel_reio}. In general, any emitted radiation is connected to the mass of the source halo ($M_{\rm h}$) as well as its accretion rate ($\dot{M_{\rm h}}$). We will first define the star-formation rate before providing details about the spectral distribution of photons. For the latter, we assume independent parametrisations for the three spectral bands of the X-ray regime, the UV regime beyond the Lyman-limit frequency, and the UV regime between the Lyman-$\alpha$ and Lyman-limit frequencies.

\subsubsection{Star formation rate}
We relate the star formation rate (SFR) of sources with the halo accretion rate via the relation
\begin{equation}
  \dot{M_{*}} = f_*(M_{h}) \dot{M_{h}},
  \label{eq:f_star_definition}
\end{equation}
where the star-formation efficiency (SFE) function is given by the double power-law
\begin{equation}
  f_*(M_{h}) = \frac{2(\Omega_{b}/\Omega_{m})f_{*,0}}{(M_{h}/M_{p})^{\gamma_{1}}+(M_{h}/M_{p})^{\gamma_{2}}}\times S(M_{h}),
  \label{eq:f_star_functional_form}
\end{equation}
with the additional low-mass term
\begin{equation}
  S(M_{h}) = [1+(M_{t}/M_{h})^{\gamma_{3}}]^{\gamma_{4}} .
  \label{eq:S(M)}
\end{equation}
The shape of $f_*$ is motivated by abundance matching results between observed galaxy luminosity and simulated halo mass functions at $z\simeq 0-10$ \citep{Behroozi_2013_fstar}. The term $S(M_{h})$ may either act as a truncation or an enhancement term of the SFE at small halo masses (depending on the parameter choices of $\gamma_3$ and $\gamma_4$) accounting for our current ignorance of the star formation process at the smallest scales. 

At high redshifts, the mass growth of haloes is known to be well described by an exponential function of the form
\begin{equation}
    M_{h}(z) = M_{h}(z_{0})\exp{[-\alpha (z-z_{0})]},
  \label{eq:EXP MAR}
\end{equation}
with $\alpha=0.79$ \citep{MAR_alpha_079}. We use this exponential model (EXP) in this paper. In the left-most panel of Fig.~\ref{fig:Profiles}, we illustrate the agreement of the exponential halo growth model (black line) with simulations from \citep{B20_Simulation}. Other methods exist to quantify the growth of haloes, relying on Abundance Matching or on the extended Press-Schechter formalism \citep{HaloModel_Paper}. They are also implemented in {\tt BEoRN}. 

\subsubsection{Lyman-$\alpha$ emissivity}
The emissivity $\epsilon_{\alpha}$ [s$^{-1}$Hz$^{-1}$($M_{\odot}$/yr)$^{-1}$] of UV photons  between the Lyman-$\alpha$ ($\nu_{\alpha}$) and Lyman-limit ($\nu_{\rm LL}$) frequencies is given by
\begin{equation}
  \epsilon_{\alpha}(\nu) = \frac{N_{\alpha}}{m_p  }I_{\alpha}(\nu),
  \label{eq:eps_al}
\end{equation}
where $I_{\alpha}(\nu)\propto \nu^{-\alpha_{s}}$ in [$\rm Hz^{-1}$] is the normalised spectral shape function, $m_{\rm p}$ is the proton mass (expressed in $M_{\odot}$), and $N_{\rm \alpha}$ the number of photons per baryons in stars emitted between $\nu_{\alpha}$ and $\nu_{\rm LL}$. Typical values are $N_{\rm \alpha}=9690,\, 4800$, respectively for Population II, and Population III stars \citep{Barkana_2005}. 

\subsubsection{X-ray emissivity}
The emissivity $\epsilon_{\rm X}$ [s$^{-1}$Hz$^{-1}$($M_{\odot}$/yr)$^{-1}$] of X-ray photons is given by
\begin{equation}
  \epsilon_{\rm X}(\nu) = c_{\rm X}\cdot f_{\rm X} \frac{I_{\rm X}(\nu)}{\nu \cdot h_{\rm P}}
  \label{eq:eps_xray}
\end{equation}
where the spectrum is again assumed to have a power-law shape $I_{\rm X}(\nu)\propto \nu^{-\alpha_{x}}$ [$\rm Hz^{-1}$]. The additional normalisation factor $c_{\rm X}$ is obtained from observation by integrating X-ray fluxes of local sources \citep{Mineo_2012_xrays,Fragos_2013, Cited_By_Jordan_Edges} in the energy range ($E_{\rm min, norm}$, $E_{\rm max, norm}$). The values of $c_{\rm X}$, $E_{\rm min, norm}$, and $E_{\rm max, norm}$ are left free, and depend on the X-ray sources, which are usually assumed to be either high-mass X-ray binaries (HMXB) or quasars (QSO). A standard value used in the literature is $c_{\rm X}=10^{40.5}$ erg~s$^{-1}$yr~$M_{\odot}^{-1}$ in the energy range $E_{\rm min, norm}=500$ eV, $E_{\rm max, norm}=2$ keV \citep{Park_18}, but $c_{\rm X}$ remains largely unknown for high redshift sources. Lastly, to account for the absorption of soft X-ray photons by the host galaxy \citep{Das_2017}, we only allow the X-ray spectrum to extend over an energy range [$E_{\rm min}$, $E_{\rm max}$]. Reducing the value of $E_{\rm min}$ can significantly increase the heating of the gas by allowing for more soft X-ray photons to reach the IGM. 
\begin{figure*}
    \centering
    \includegraphics[width =1\textwidth,trim=0cm 0.0cm 0cm 0.0cm,clip]{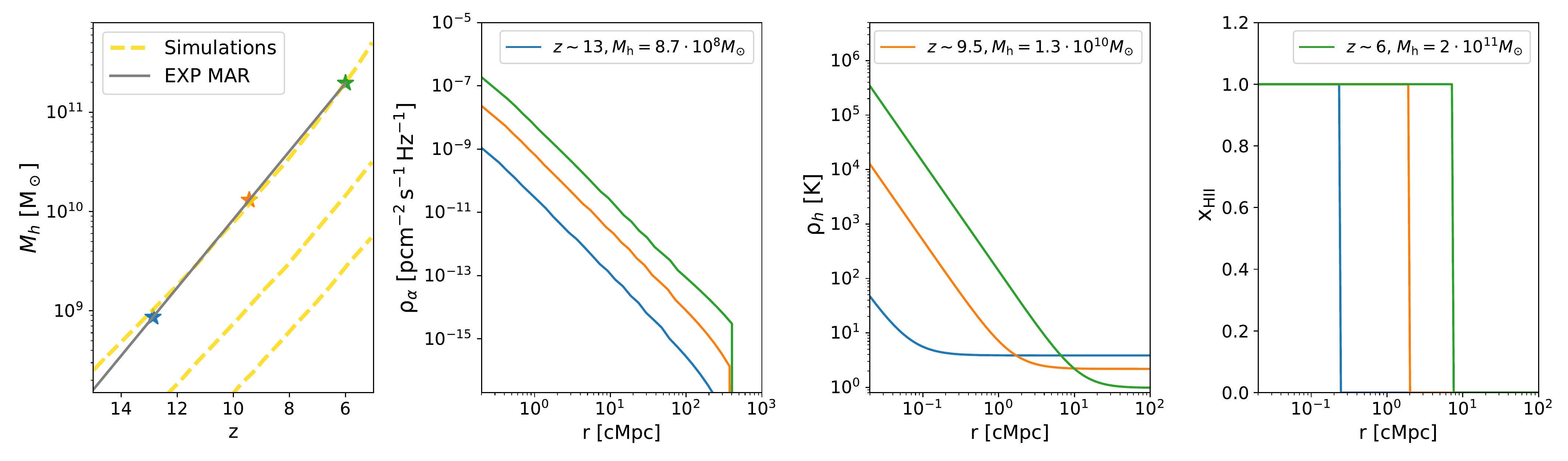}
    \caption{
    Halo mass growth and profiles for different masses and redshifts. \textit{First panel:} The grey curve depicts the halo mass accretion history as predicted by the EXP MAR model (Eq.~\ref{eq:EXP MAR}). We compare it to the yellow curves, obtained from simulation \citep{B20_Simulation}.  \textit{Second panel:} Profiles for the Lyman-$\alpha$ coupling coefficient $x_{\rm \alpha}$ (Eq.~\ref{eq:xal} and \ref{eq:rho_alpha}), as a function of comoving distance $r$. \textit{Third panel:} Kinetic temperature profiles $T_{\rm k}$ (Eq.~\ref{eq:heat equation}). \textit{Fourth panel :} Ionisation fraction profile $x_{\rm HII}$ (Eq.~\ref{eq:Stromgren Sphere}).}
    \label{fig:Profiles}
\end{figure*}

\subsubsection{Ionising photons}
We parametrise the galactic ionising radiation differently than the Lyman-$\alpha$ and X-ray components. Instead of a frequency-dependent spectrum, we only account for the total number of photons with energy larger than 13.6 eV emitted per unit time ($\dot{N}_{\rm ion}$): 
\begin{equation}
  \dot{N}_{\rm ion}(M_{h}) = f_{\rm esc}(M_{h})f_{*}(M_{h})\dot{M}_{h} \times \frac{N_{\rm ion}}{m_{\rm p}},
  \label{eq:Niondot}
\end{equation}
where $N_{\rm  ion}$ stands for the number of ionising photons per baryons in stars, and $f_{\rm esc}$ is the escape fraction of ionising photons. The escape fraction is assumed to scale as a power law
\begin{equation}
  f_{\rm esc}(M_{\rm h}) = f_{\rm esc, 10}(M_{h})\left( \frac{10^{10}M_{\odot}}{M_{h}}\right)^{\alpha_{\rm esc}}.
  \label{eq:fesc}
\end{equation}
The function is truncated so that $f_{\rm esc}\leq1$. Note that radiative-hydrodynamical simulations point towards a decreasing $f_{\rm esc}$ with halo mass, which is explained by the low column density of gas in small mass galaxies \citep{f_esc_Paardekooper_2015,f_esc_hydro_Taysun,f_esc_Simu_Tessyier}.

Note that to calculate the emission rate of photons in the Lyman-$\alpha$ and X-ray bands (Eq.~\ref{eq:eps_al} and \ref{eq:eps_xray}, respectively), the SFR obtained from the halo mass via Eq.~\ref{eq:f_star_definition} and \ref{eq:EXP MAR} needs to be multiplied. On the other hand, the ionizing photon rate is directly provided by Eq.~\ref{eq:Niondot}.

\subsection{Radial profiles}
\label{sec2:5}
As mentioned in the beginning, one of the core ingredients of the \beorn framework are the spherically symmetric profiles describing the Lyman-$\alpha$ coupling ($x_{\rm \alpha}$), the gas temperature ($T_{\rm k}$), and ionization process ($x_{\rm HII}$) around sources. In this section, we provide an overview of the equations governing the evolution of these profiles. More detailed descriptions of the calculations can be found in \citet{HaloModel_Paper, HaloModel_reio}.

\subsubsection{Lyman-$\alpha$ flux profiles}
The interactions of the Lyman-$\alpha$ photons with the neutral hydrogen atoms of the IGM leads to a coupling of the spin temperature to the kinetic temperature of the gas \citep{Wouthuysen1952aaa,Field1958aaa}. This coupling causes an absorption trough in the global signal prior to the epoch of reionization. The value of the flux at a physical distance $r$ from a source with mass $M_{\mathrm{h}}$ at time $z$ is computed by tracing back the photons emitted at $r=0$ at an earlier time $z'$, when the source had a smaller mass $M_{h}(z')$. This resulting flux profile is given by
\begin{equation}
  \rho_{\alpha}(r|M_{h},z) = \frac{1}{4\pi r^{2}} \underset{n=2}{\overset{n_{m}}{\sum}}f_{n} \epsilon_{\alpha}(\nu_{\rm n}') f_*(M_{h}(z'))\dot{M}_{h}(z')
  \label{eq:rho_alpha}
\end{equation}
with $\nu_{\rm n}' = \nu_{\rm n}(1+z')/(1+z)$, where $\nu_{\rm n}$ is the frequency of the Lyn resonance, and where $f_{n}$ are the recycling fractions assuming the truncation $n_{m} = 23$ \citep{Pritchard:2005an}. The look-back redshift $z'$ is obtained
by numerically inverting the comoving distance relation
\begin{equation}
  r= \frac{1}{1+z}\int_{z}^{z'} \frac{c}{H(z'')} \,dz'',
  \label{eq:rcom inverted}
\end{equation}
where $H(z)$ refers to the Hubble parameter.

In the second panel of Fig.~\ref{fig:Profiles} we show examples of Lyman-$\alpha$ flux profiles for different halo masses and redshifts (as indicated by the coloured stars in the left-most panel). The underlying source model assumes a flat $f_*=0.1$, leading to an exponential growth in the SFR. The profiles decrease as $1/r^{2}$ close to the source steepening gradually  towards larger radii due to the look-back effect and the lower star formation rate in the past. 
{The steep cutoff at a few hundred Mpc is due to the horizon beyond which the source becomes invisible}.


\subsubsection{Temperature profiles}
The heating of the IGM around sources is caused by X-ray radiation which interacts with the surrounding gas cells. The relevant X-ray emission profiles are given by
\begin{multline}\label{rhoxray}
\rho_{\rm xray}(r|M,z)= \frac{1}{4\pi r^2}\sum_{i}f_if_{X,h}\\
\times\int_{\nu_{\rm th}^i}^{\infty} d\nu (\nu-\nu_{\rm th}^i)h_P\sigma_i(\nu)\varepsilon_{X}(\nu'){\rm e}^{\tau_{\nu'}}f_* {\dot M}_{h}(z').
\end{multline}
where $i=\{\rm H,\rm He\}$ and $\nu_{\rm th}^i=\{13.6, 26.5\}$ eV. $\rho_{\rm xray}$ has units of $[\rm eV/s]$. Note that $r$ here again is a physical distance. The mean optical depth of the IGM ($\tau_{\nu'}$) is computed according to Eq.~(26) of \citet{HaloModel_Paper}. $h_P$ is the Planck constant. $f_{X,h}$ is the amount of heat deposited by secondary electrons. We assume $f_{\rm X,h} = \bar{x}_{\rm e}^{0.225}$ \citep{Shull_vanSteenberg_1985}, where $\bar{x}_{\rm e}$ is the mean free electron fraction in the neutral medium. We compute $\bar{x}_{\rm e}$ according to eq.~9, 10 and 12 in \citet{Mirocha_Decoding_2014}. 

The temperature profile $\rho_{h}$ (in $[\rm K]$) of the neutral medium around a source is then obtained by solving the following differential equation: 

\begin{equation}
  \frac{3}{2} \frac{d\rho_{h}(r|M_{h},z)}{dz} = \frac{3\rho_{h}(r|M_{h},z)}{(1+z)}-\frac{\rho_{\rm xray}(r|M_{h},z)}{k_{\rm B}(1+z)H(z)}.
  \label{eq:heat equation}
\end{equation}
We show example $\rho_{h}$ 
profiles in the third panel of Fig.~\ref{fig:Profiles} that are obtained by solving Eq.~\ref{eq:heat equation} assuming constant adiabatic initial conditions. These profiles decrease as $1/r^{2}$ and reach the adiabatic cooling plateau at large radii, representing regions far away from the source where baryons have not been heated by X-ray photons yet.


\begin{figure*}
    \centering
    \includegraphics[width =1.01\textwidth,trim=4.5cm 0cm 4.5cm 0cm,clip]{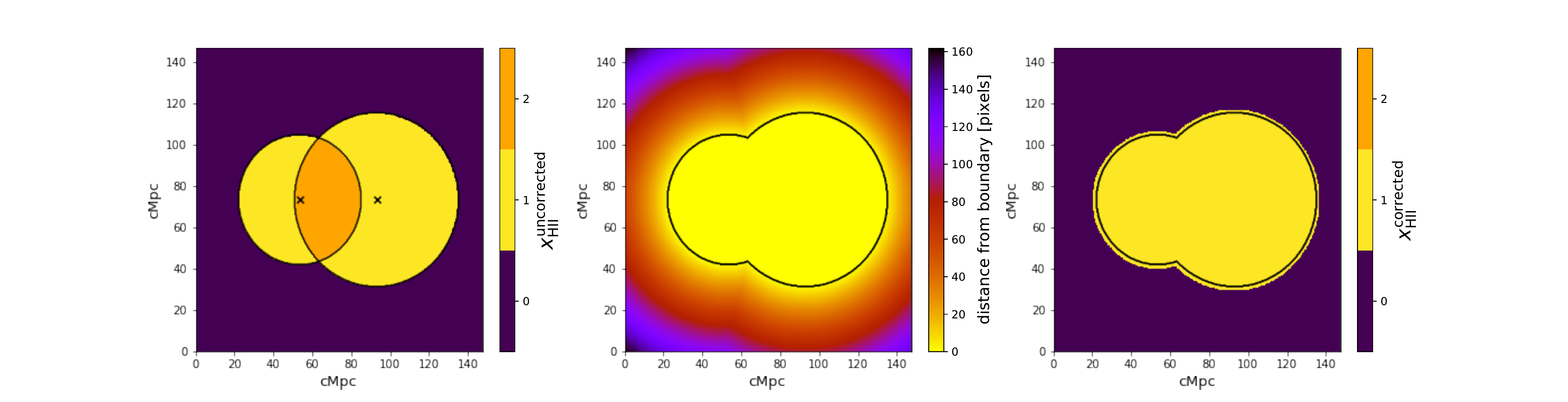}
    \caption{
    A toy example to illustrate how we correct for the overlap of ionised bubbles. \textit{Left:} Two overlapping ionised bubbles before correcting for overlap. The black crosses indicate the positions of the two sources. Black contours correspond to the external border of the two ionised bubbles. 
    The purple background corresponds to neutral hydrogen ($x_{\rm HII} = 0$) and the yellow and orange regions to ionised pixels ($x_{\rm HII} \geq 1$). The bubbles overlap and lead to ``overionised'' (orange) pixels, marked with $x_{\rm HII} = 2$.  \textit{Middle:} We flag the background neutral pixels with their distance from the boundary of the ionised region.  \textit{Right:} The two ionised bubbles, after correcting for overlap. We have redistributed the excess ionisation fraction to a set of pixels closest to the boundary of the ionised region, which grows subsequently. We plot the same black contour as in the left panel.}
    \label{fig:BubbleSpreading}
\end{figure*}

\subsubsection{Ionisation fraction profiles}
The co-moving volume $V$ of an HII region around  a source emitting ionising photons at a rate $\dot{N}_{\rm ion}(t)$ [$\rm s^{-1}$] evolves according to:
\begin{equation}
  \frac{dV}{dt} = \frac{\dot{N}_{\rm ion}(t)}{\bar{n}_{H}^{0}} -\alpha_{B} \frac{C}{a^{3}}\bar{n}_{H}^{0}V,
  \label{eq:Stromgren Sphere}
\end{equation}
with $\alpha_{B}$ the case-B recombination coefficient, $C$ the clumping factor, $a$ the scale factor,  $\bar{n}_{H}^{0}$ the mean co-moving number density of hydrogen. The ionisation fraction profile $x_{\rm HII}$ is then given by a Heaviside step function:
\begin{equation}
  x_{\rm HII}(r|M,z)= \theta_H\left[R_{b}(M,z)-r\right],
  \label{eq:ion frac profile}
\end{equation}
with the co-moving bubble size $R_{b}$ obtained from the volume $V$ via $V = 4\pi R_{\rm b}^{3}/3$.
Ionisation profiles are plotted in the rightmost panel of Fig.~\ref{fig:Profiles}. The sharp ionisation front of the bubble expands as the source grows and emits photons. Note that in our framework the clumpiness of the IGM is controlled by the parameter $C$. In this work, we will assume $C=1$ for simplicity.


\subsection{Constructing three dimensional maps}\label{sec2:6}

The aim of \beorn is to generate 3D simulation volumes of the IGM during the epoch of reionisation and cosmic dawn, which can be observed with the 21-cm signal. To create 3D maps of the 21-cm signal or $dT_{b}(z,\vec{x})$, temperature, Lyman-$\alpha$ profiles, and ionised bubbles are painted on a grid around photon sources. To paint profiles on a grid $\mathcal{G}$, 3D convolution is used, which can be defined as follows:
\begin{eqnarray}
\mathcal{G}(\vec{x}) = \sum_{\vec{x}_{h}} \int \delta^{3D}_{\vec{x}_h} (\vec{x}) \mathcal{P}(\vec{y}-\vec{x}) d\vec{x} \ ,
\label{eq:paint_convolve}
\end{eqnarray}
where $\delta^{3D}$ is the 3D Kronecker delta function that equals one at the location of halo $\vec{x}_h$ and zero everywhere else. As the sources are assumed to be present in dark matter haloes in a spherically symmetric environment, $\mathcal{P}$ represents 3D grids created from the radial profiles (see Sec.~\ref{sec2:5}).

We use an optimised version of the fast Fourier transform (FFT) based 3D convolution implemented in the {\tt astropy}\footnote{\url{https://docs.astropy.org}} package. However, the computational cost of performing this step for every halo found in the simulation becomes prohibitively high. For example, simulation volumes at redshift $z=6$ with a box length $\gtrsim$100 cMpc have more than 10 million haloes if we resolve haloe-masses down to $\sim10^9 ~M_\odot$\footnote{Note that the number of haloes increases both with decreasing redshift and with the resolution limit of the simulation.}. To reduce computing time, we bin halo masses and assume that every halo in a given mass bin shares the same profiles.



Let us assume we want to create $dT_{b}$ coeval boxes for a range of redshifts ($z_{j}$). We first initialise an array of halo masses $M_{\rm bin}$, of size $N_{\rm bin}$, spanning from $M_{\rm min}$ to $M_{\rm max}$ (assuming logarithmically spaced binning), corresponding to halo masses at the final redshift $z_{N_{\rm z}}$ of the simulation. Then, we compute the mass accretion history of each mass element, backwards in time, according to the chosen MAR model (Eq.~\ref{eq:EXP MAR}). We end up with a 2D array of masses ($M_{\rm i,j}$) with $i=0..N_{\rm bin}$, and $j=0..N_{\rm z}$. Finally, we compute three profiles, denoted as $\mathcal{P}$, for each mass bin. These profiles are obtained by evaluating Eqs. \ref{eq:rho_alpha}, \ref{eq:heat equation} and \ref{eq:ion frac profile}, starting from $z_{0}$ down to $z_{N_{\rm z}}$, while tracking the evolution of the SFR using Eq.~\ref{eq:f_star_definition}. As a result, we obtain $3\times N_{z}$ profiles per mass bin, leading to a total of $3\times N_{\rm bin}\times N_{z}$ profiles.

We can visualise this process through Fig.~\ref{fig:Profiles}. The grey line in the leftmost panel shows the mass evolution of one halo mass bin, while the other panel shows the subsequent evolution of the profiles. To construct a map at a given redshift, we loop over the $N_{\rm bin}$ mass bins and simultaneously paint all profiles in this halo mass range. The timing of this step scales with $N_{\rm bin}$. The input parameter $N_{\rm bin}$, $M_{\rm min}$, and $M_{\rm max}$ are left to the user and should be chosen such that the array of binned masses covers all halo masses included in the input halo catalogue at each redshift $z_j$.

We note that our approach makes two approximations: (1) we treat haloes within the same mass bin as identical, even though they have slightly different masses and hence different profiles, and (2) we assume that haloes reside at the centre of the grid cell in which their true centre is. These approximations disappear in the limit of infinite grid cells and halo mass bins. To determine a suitable number of mass bins for converged power spectra, we perform convergence checks. In Appendix~\ref{Appendix:Convergence}, we compare the 21-cm power spectra obtained when varying the number of grid cells $N_{\rm cell}$ and mass bins $N_{\rm bin}$.

\subsubsection{Overlap of Lyman-$\alpha$ and temperature profiles}
Both the Lyman-$\alpha$ and temperature profiles typically extend far beyond the corresponding halo boundaries. As a consequence, many regions of the IGM will be inside the area of influence of two or several profiles from different neighbouring sources. For the case of the Lyman-$\alpha$ and X-ray emission, it is obvious that multiple contributions from different sources can be added up to obtain a total flux per gas cell.

Regarding the heating of the gas, the situation is more complicated as the temperature is obtained via a differential equation (Eq.~\ref{eq:heat equation}). Following \cite{HaloModel_Paper}, we separate the temperature field $T_{\rm k}(x,z)$ into a heating term $T_{\rm h}(x,z)$, and a primordial component $T_{\rm p}(x,z)$, such that $T_{\rm k}=T_{\rm p}+T_{\rm h}$. The primordial component is sourced by the matter fluctuations and evolves according to
\begin{equation}
T_{\rm p}(x,z) = T_{\rm ad}(z)(1+\delta_{b}(x))^{2/3},
\end{equation}
where $T_{\rm ad}\propto (1+z)^2$ corresponds to the adiabatic gas temperature that decoupled from the CMB temperature at $z \sim 135$. 

The heating term $T_{\rm h}$ is obtained by summing up all radial profile components $\rho_{\rm heat}$ that affect the IGM at a given position. This summation procedure becomes possible uniquely due to the simple form of Eq.~\ref{eq:heat equation}. If we consider, for example, two separate sources with X-ray flux profiles ($\rho_{\rm xray,1}$,  $\rho_{\rm xray,2}$), leading to temperature profiles ($\rho_{\rm h,1}$, $\rho_{\rm h,2}$), the solution of Eq.~\ref{eq:heat equation} for a source with x-ray profile $\rho_{\rm xray,1}$+$\rho_{\rm xray,2}$ is $\rho_{\rm h,1}$ + $\rho_{\rm h,2}$. Hence, we are allowed to add up on a grid the heating profiles $\rho_{\rm h}$ from different sources. The final gas temperature maps $T_{\rm k}(x,z)$ are obtained by adding $T_{\rm p}$ and $T_{\rm h}$.

Note that the additive nature of temperature is tied to the simplicity, separability, and linearity of Eq.~\ref{eq:heat equation}, a differential equation. However, this characteristic is lost when considering a spatially varying $x_e$ component, which, in principle, influences the temperature calculation \citep[see e.g., eq.~11 of][]{21cmFAST_2010}. Nevertheless, we have verified that the additional term associated with $x_e$ remains negligible for all individual profiles. Hence, it can be safely disregarded.  


\begin{figure*}
     \centering
     \includegraphics[width =1\textwidth,clip]{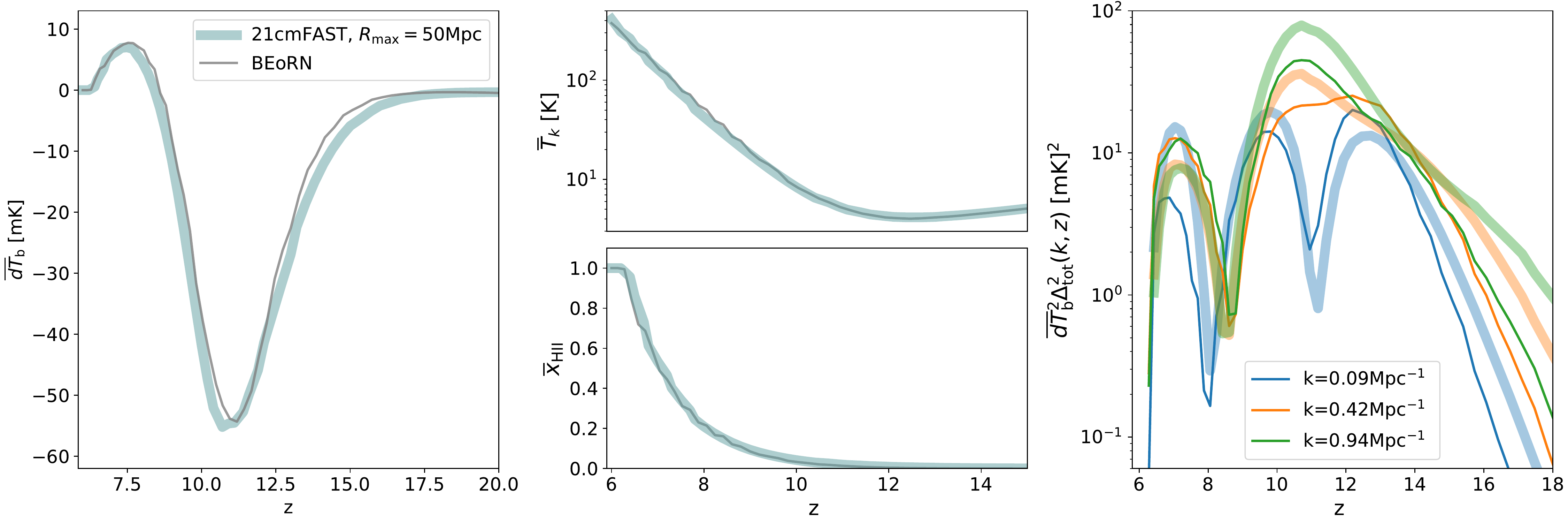}
     \caption{
     Comparison between {\tt 21cmFAST} (thick lines in the four panels) and {\tt BEoRN} (thin lines). RSD are not included. \textit{Left:} Global differential brightness temperature $dT_{\rm b}$.  \textit{Upper middle:} Mean kinetic temperature of the gas $T_{\rm k}$. \textit{Lower middle:} Mean ionisation fraction history $x_{\rm HII}$. \textit{Right:} Dimensionless power spectrum of the $dT_{\rm b}$ field, shown as a function of redshift $z$, at three different scales, k = 0.09, 0.42, 0.094 Mpc$^{-1}$ respectively in blue, orange and green.}
     \label{fig:21cmFAST_BEoRN_Comparison}
\end{figure*}

\subsubsection{Overlap of ionisation bubbles}\label{sec2:7}
In contrast to the Lyman-$\alpha$ and temperature profiles, the ionising bubbles are not additive. Any gas cell can only be ionized once. This means that $x_{\rm HII}$ has to stay between 0 and 1.

The painting process described by Eq.~\ref{eq:paint_convolve} will naturally add up profiles giving rise to pixels with $x_{\rm HII}>1$. In order to guarantee photon conservation, we have to take care of these "overionised" cells and distribute the excess ionisation fraction in a consistent way. Fig.~\ref{fig:BubbleSpreading} shows our approach for handling the overlap of ionised bubbles. In the illustrated example, we consider a system with two distinct sources that produce two overlapping bubbles. The left panel shows the resulting system after we paint the bubbles on a grid but before correcting for the overlap. The orange region indicates where the bubbles overlap and is marked with an unphysical ionising fraction of $x_{\rm HII}=2$. We note that this problem is similar in nature to the photon non-conservation issue discussed in e.g. \citet{Choudhury_2018}.

Realistic simulations will contain a large number of overlapping regions. We identify all these regions using the method implemented in {\tt skimage}\footnote{\url{https://scikit-image.org/}} package to label connected components \citep{fiorio1996two_skimagelabel,wu2005optimizing_skimagelabel}. If these connected regions contain "overionised" cells, we use the distance transform algorithm in the {\tt scipy}\footnote{\url{https://scipy.org/}} package that gives the distance from the nearest boundary in terms of number of pixels. As we need to distribute the excess ionisation in the neutral pixels, we run this algorithm to get distance $d=0$ inside ionised regions and $d>0$ outside. This process is illustrated in the middle panel of Fig.~\ref{fig:BubbleSpreading}. We distribute the excess ionisation among the neutral pixels by weighting over the distances. 
The excess ionisation is distributed only after pixels at a shorter distance have been fully ionised.
In the right panel of Fig.~\ref{fig:BubbleSpreading}, we show the system after redistributing the excess ionisation fraction to the border of the ionised region. As a result, the overall ionised island expanded proportionately to the volume of the overlapping region.

At early epochs of reionisation, the procedure of redistributing photons is time-consuming as there is a plethora of very small ionised regions that only contain a few ionised pixels. In order to speed up the code, we introduce a parameter $N_{\rm cell, th}$, which groups all small ionised islands containing fewer than $N_{\rm cell, th}$ pixels together. In Appendix~\ref{Appendix:overlapp} we show the  effects of increasing $N_{\rm cell, th}$ on the $x_{\rm HII}$ power spectrum. By choosing an appropriate value for $N_{\rm cell, th}$, we can significantly reduce the computing time of the code without affecting the power spectrum over the scales relevant for 21-cm interferometers. Unless stated otherwise, we use the default value $N_{\rm cell, th} = 160\times (N_{\rm cell}/256^{3})$, with $N_{\rm cell}$ the total number of pixels.


\begin{figure*}
     \centering
     \includegraphics[width =1\textwidth,clip]{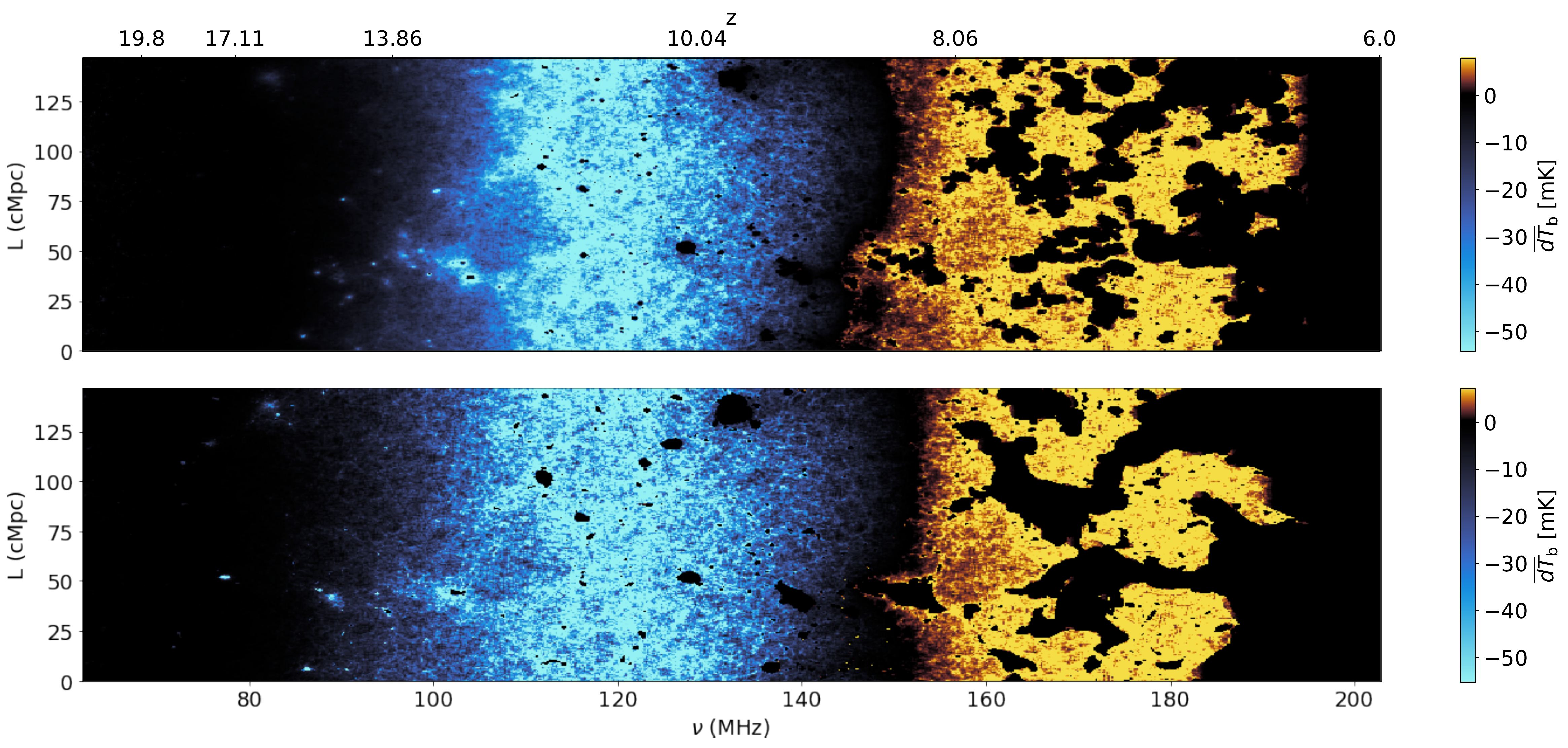}
     \caption{Comparison of lightcone maps for the differential brightness temperature from \beorn (top) and {\tt 21cmFAST} (bottom). RSD are not included. Blue and yellow coloured pixels correspond to regions dominated by absorption ($dT_{\rm b}<0$) and emission ($dT_{\rm b}>0$), respectively. The black patches represent the ionised bubbles ($dT_{\rm b}=0$). As illustrated, these bubbles merge and fill the universe by the time it reaches $z\sim 6$.}
     \label{fig:lightcones_FAST_BEORN}
\end{figure*}

\section{Validation of our framework}
\label{sec3}
As a next step, we validate {\tt BEoRN} by comparing it to the widely used semi-numerical simulation code {\tt 21cmFAST}\footnote{This code is under constant development and therefore multiple versions are available. In this work, we used the C version that can be found at \url{https://github.com/andreimesinger/21cmFAST}. The python-wrapped version of {\tt 21cmFAST} \citep{murray202021cmfast} is currently being upgraded with more accurate gravity evolution calculation (Andrei Mesinger, private communication) and we plan to provide an interface in \beorn~to this version.} \citep{21cmFAST_2010}. In {\tt 21cmFAST}, the matter density field is evolved perturbatively according to the Zel'Dovich approximation. Ionisation maps are created based on the excursion set formalism \citep{Exc_Set_Furla_2004_TheGrowthHIIRegions}. This approach is fundamentally different from {\tt BEoRN} since sources are not resolved individually. Instead, ionisation, Lyman-$\alpha$ coupling, and heating rates are computed from the collapsed fraction field $f_{\rm coll}$, obtained by integrating the Press-Schechter sub-halo mass function. Furthermore, in {\tt 21cmFAST} the stellar-to-halo relation and the star formation rate are computed differently than in {\tt BEoRN} (see eq.~2-3 in \citealt{Park_18}), the spectral energy distribution of Lyman-$\alpha$ photons is obtained assuming the piece-wise power-law interpolation from \cite{Barkana_2005}, and inhomogeneous recombinations are implemented. It is therefore difficult to find an exact mapping between the {\tt 21cmFAST} and {\tt BEoRN} source parameters, and a certain amount of recalibration is required for a comparison. 

The comparison is performed by (i) running {\tt 21cmFAST} for a generic set of astrophysical and cosmological parameters, (ii) extracting the halo catalogues and density field produced by {\tt 21cmFAST} as input files for {\tt BEoRN}, (iii) re-calibrating, if needed, the intensity of Lyman-$\alpha$, X-ray, and ionising radiation of {\tt BEoRN} (controlled by $N_{\rm \alpha}$, $f_{\rm X}$, and $N_{\rm ion}$), in order to match the relevant global quantities between the two codes (i.e the temperature $T_{\rm k}$, the ionisation history $x_{\rm HII}$, and the 21-cm global signal $dT_{\rm b}$). Once the global quantities agree, we compare the power spectra of the $dT_{\rm b}$ fields at different redshifts and $k$-modes. The re-calibration of the global parameters is discussed in Sec.~\ref{sec3:2}.



We run {\tt 21cmFAST} assuming a coarse and fine grid resolution ({\tt DIM}, {\tt HII\_DIM}) of 300$^{3}$ and 900$^{3}$ cells, in a box with side length equal to 147 cMpc. The parameter controlling the maximum size of ionised bubbles (interpreted as being the mean free path of ionising photons) is set to the default value of $R_{\rm max}=50 \rm \, Mpc$. The stellar-to-halo-mass relation in {\tt 21cmFAST} is parameterised as a power law, followed by an exponential cutoff towards small scales (see eq.~2 in \citealt{Park_18}). We set its normalisation to $f_{\ast,10}=0.02$, the power law index to $\alpha_{\ast}=0.5$, and the turnover mass to $M_{\rm turn} = 10^{9}M_{\odot}$. We furthermore assume a UV photon number of $N_{\gamma,uv} = 5000$ as well as a flat escape fraction with normalisation $f_{\rm esc,10} = 0.25$. The X-ray spectral index is set to $\alpha_{X}=1.5$, and the X-ray flux normalisation to $L_{X} = 3.4 \times 10^{40 } \rm{erg .s^{-1}. M_{\odot}^{-1}. yr}$, in the energy range $E_{\rm min}=500 \rm eV$, $E_{\rm max} = 2 \rm keV$. Finally, we turn off redshift space distortions for the comparison.

Next to the default {\tt 21cmFAST} simulation specified above, we rerun {\tt 21cmFAST}, assuming the exact same setup (with identical random seed), this time turning on the embedded halo algorithm (${\tt USE HALOS = 1}$). The created halo catalogues as well as the corresponding dark matter density fields are then used as input files for {\tt BEoRN}. This allows for a comparison based on the same realisation of the density and source fields minimising the impact of cosmic variance.



\subsection{Global quantities}\label{sec3:2}
To ensure consistency of the global quantities, such as the mean IGM temperature ($\overline{T}_{\rm k}$), the ionization history ($\overline{x}_{\rm HII}$), and the sky-averaged 21-cm signal ($\overline{dT}_{\rm b}$) between the two codes, we fit the different stellar-to-halo relation ($f_{\ast}$) and escape fraction ($f_{\rm esc}$) to the model of {\tt 21cmFAST} and recalibrate a few selected parameters. The best fit for $f_{\ast}$(M) is obtained assuming the values $f_{\ast,0} = 0.02$, $\gamma_{1} = -0.5$, $\gamma_{2} = -0.5$, $\gamma_{3} = 1.4$, $\gamma_{4} = -4$, $M_{p} = 10^{10}$ M${_\odot}$, and $M_{t} = 4\times10^{8}$ M${_\odot}$ in Eq.~(\ref{eq:f_star_functional_form}). The escape fraction of ionising photons and the X-ray flux are parameterised in the same way in both codes. We hence set $f_{\rm esc,10} = 0.25$ and $\alpha_{\rm esc}=0$ in Eq.~(\ref{eq:fesc}) as well as $c_{\rm X}=3.4 \times 10^{40 } \rm{erg .s^{-1}. M_{\odot}^{-1}. yr}$, $\alpha_{X}=1.5$, $E_{\rm min} = E_{\rm min, norm} = 500 \rm eV $ and $E_{\rm max} = E_{\rm max, norm} = 2 \rm keV $ in Eq.~(\ref{eq:eps_xray}).

Finally, we assume a flat Lyman-$\alpha$ spectral index $\alpha_{\rm s}=0$ in Eq.~(\ref{eq:eps_al}), instead of the piece-wise power-law used in {\tt 21cmFAST}. We then fine-tune $N_{\rm \alpha}$, $f_{\rm X}$, and $N_{\rm ion}$ until we exactly match the  global quantities $\overline{T}_{\rm k}(z)$, $\overline{x}_{\rm HII}(z)$, and $\overline{dT}_{\rm b}(z)$,.

In Fig.~\ref{fig:21cmFAST_BEoRN_Comparison}, we present the outcome of the aforementioned process. The {\tt 21cmFAST} predictions are illustrated as thick lines, while the {\tt BEoRN} data are shown as thin lines. The bottom-central panel of Fig.~\ref{fig:21cmFAST_BEoRN_Comparison} shows the ionisation histories $x_{\rm HII}(z)$, which agree well between the two codes for the same number of ionising photons per baryon in stars: $N_{\rm ion}= N_{\gamma,uv}=5000$. 
To match the mean kinetic gas temperature (shown in the top-central panel of Fig.~\ref{fig:21cmFAST_BEoRN_Comparison}), we set the X-ray normalisation to $f_{\rm X}=0.8$. We speculate that this mismatch is related to the different treatment of $f_{\rm X, h}$ in the two codes. To achieve agreement in the global signal between redshifts $11<z<18$, we adjust the number of Lyman-$\alpha$ photons per stellar baryon $N_{\alpha}$. The best match between the two curves was obtained for $N_{\alpha}=3000$. Note that the global signal predicted by {\tt BEoRN} is steeper during Lyman-$\alpha$ coupling ($13<z<18$) than the one from {\tt 21cmFAST}. We speculate that this variation is caused by differences in the star formation rate densities resulting from the collapse fraction and halo fields.



\subsection{Power spectrum and lightcone images}

After ensuring that the global quantities agree, we compare the power spectra between the two codes. The spherically averaged power spectrum $P_{\rm tot}(k)$ of the $dT_{\rm b}$ fluctuations is defined as
\begin{equation}
    \big<dT_{\rm b}(k) dT^{*}_{\rm b}(k')\big> = (2\pi)^{3}\delta^{3D}(k-k')\vert \overline{dT}_{\rm b}\vert^{2}  P_{\rm tot}(k),
    \label{eq:PS def}
\end{equation}
where $\delta^{ 3D}$ is the three dimensional Dirac delta. Throughout the paper we express the results in terms of the 'dimensionless' power spectrum $\Delta_{\rm tot}^{2}=k^{3}P_{\rm tot}(k)/2\pi^{2}$ which only depends on units of [mK]$^2$.

In the rightmost panel of Fig.~\ref{fig:21cmFAST_BEoRN_Comparison}, we plot the dimensionless power spectra as a function of redshift $z$ for three different $k$-values from {\tt BEoRN} (thin lines) and {\tt 21cmFAST} (thick lines). The results from both codes show similar features, with three distinct peaks at large scales  ($k=0.09$ $\rm Mpc^{-1}$) representing the epochs of Lyman-$\alpha$ coupling, heating, and reionisation, respectively. At smaller scales ($k=0.42$ and $0.94 \rm , Mpc^{-1}$), the Lyman-$\alpha$ and heating peaks merge, but the reionisation peak remains distinct.


The comparison between {\tt BEoRN} and {\tt 21cmFAST} reveals some similarities and differences. At very high redshifts ($z>14$), {\tt BEoRN} predicts a lower power spectrum compared to {\tt 21cmFAST}, with a difference of up to one order of magnitude, especially at small scales. During the epochs of Lyman-$\alpha$ coupling and heating ($8<z<14$), the two codes agree reasonably well, with a relative difference of about a factor of 2 or less. During reionisation ($z<8$ or $x_{\rm HII}>0.2$), {\tt BEoRN} shows significantly less power at large scales while the small scales are similar to the results of {\tt 21cmFAST}.

In Fig.~\ref{fig:lightcones_FAST_BEORN} we provide lightcone maps of $dT_{\rm b}$ field generated with {\tt BEoRN} (top) and {\tt 21cmFAST} (bottom). The slice thickness is 0.49 cMpc which corresponds to the side-length of one pixel. The blue and yellow colours show regions visible in absorption and emission, respectively. The black patches which grow in size towards lower redshifts correspond to the ionised bubbles.

Comparing the two maps of Fig.~\ref{fig:lightcones_FAST_BEORN} we see that ionizing patches form at similar locations, but differences in their morphology are visible. In general, {\tt 21cmFAST} exhibits larger and smoother bubbles compared to the predictions from {\tt BEoRN}. This general observation is in qualitative agreement with earlier findings that semi-numerical schemes predict larger and more connected ionised patches compared to simulations based on radiative transfer calculations \citep[see e.g.][]{Zahm_11_Comparison_RT_SemNum,Majumdar_14_Comparison_RT_SemNum}. It also agrees with our earlier findings of the power spectrum where at the largest scales {\tt 21cmFAST} predicted a significantly higher signal than \beorn.

In Appendix~\ref{Appendix:mfp} we study in more detail the differences between the large-scale power from {\tt 21cmFAST} and \beorn. In particular, we suggest that the difference may be caused by the different treatments of the photon mean free path. In {\tt 21cmFAST}, the mean free path of ionising photons is parameterised through $R_{\rm max}$, which limits the maximum distance photons can travel. In contrast, the mean free path in {\tt BEoRN} is an inherent feature of the code, determined by the $x_{\rm HII}$ maps and the bubble size distribution. As demonstrated by \cite{Georgiev_22}, reducing the value of $R_{\rm max}$ in {\tt 21cmFAST} leads to a shift of power from large to small scales, which is in qualitative agreement with the shift observed in Fig.~\ref{fig:21cmFAST_BEoRN_Comparison}. In Appendix~\ref{Appendix:mfp}, we furthermore compare our results with those of \citet{Davies&Furlanetto}, who developed an improved excursion-set method that accounts for the gradual absorption of ionising photons, rather than treating the mean free path as a sharp barrier (as is the case with $R_{\rm max}$). We find a better match between \beorn and the results from \citet{Davies&Furlanetto} compared to the standard $R_{\rm max}$-method implemented in {\tt 21cmFAST}.


\begin{figure}
     \centering
     \includegraphics[width =0.48\textwidth,trim=0.28cm 0.3cm 0.3cm 0.2cm,clip]{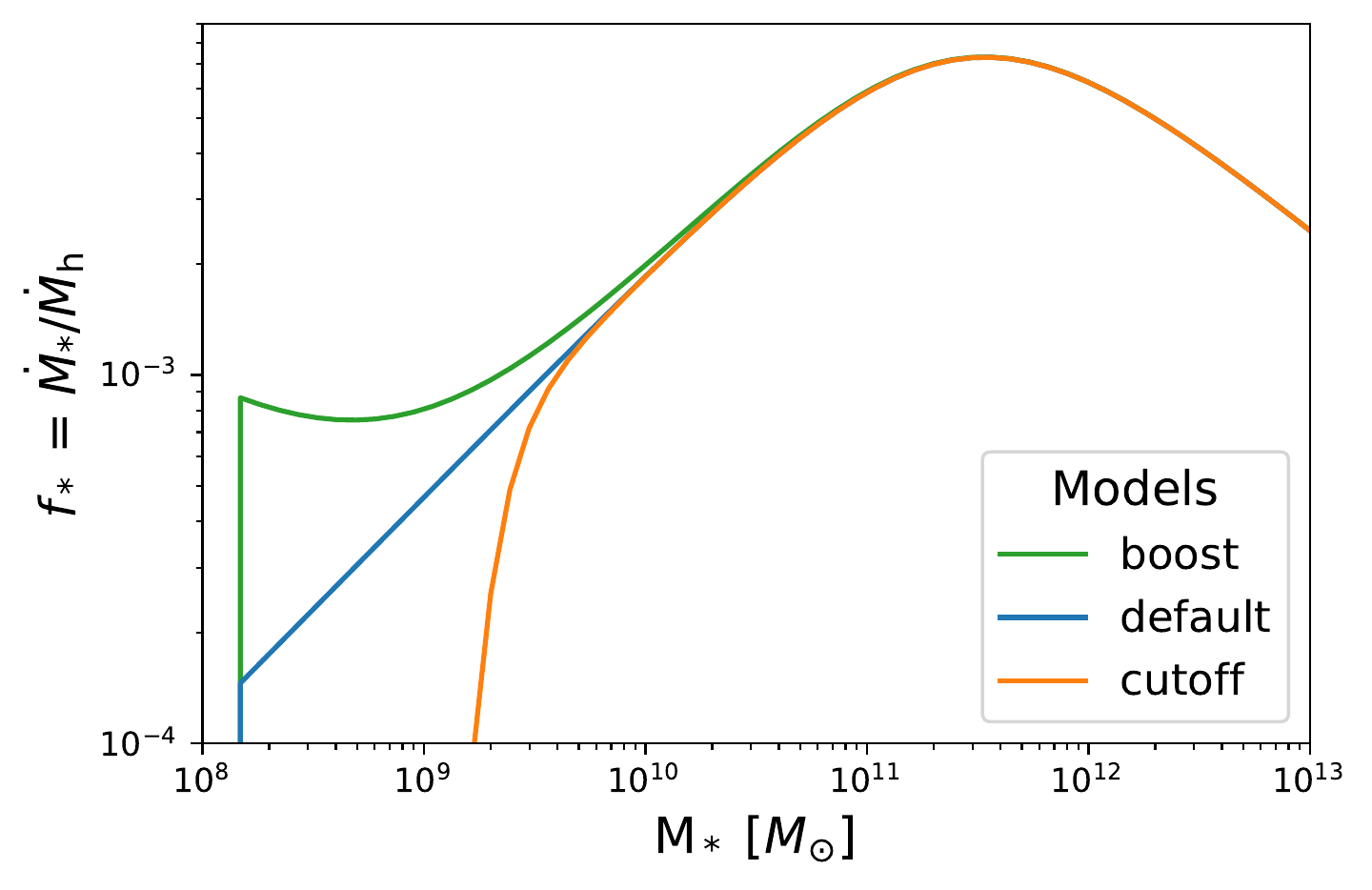}
     \caption{Stellar-to-halo mass relation for the three benchmark models characterised by a boost (green), a power-law decrease (blue, default model), and a cutoff (orange) towards small halo masses. We set $f_{*}=0$ below the minimum halo mass $M_{\rm h}=1.47\times10^{8}M_{\odot}$ of the simulation. The functional form of $f_{\ast}$ is given in Eq.~\ref{eq:f_star_functional_form} the parameter values are provided in Table~\ref{tab:3Models param}.}
     \label{fig:3models_fstar}
\end{figure}

\begin{figure*}
     \centering
     \includegraphics[width =1\textwidth,clip]{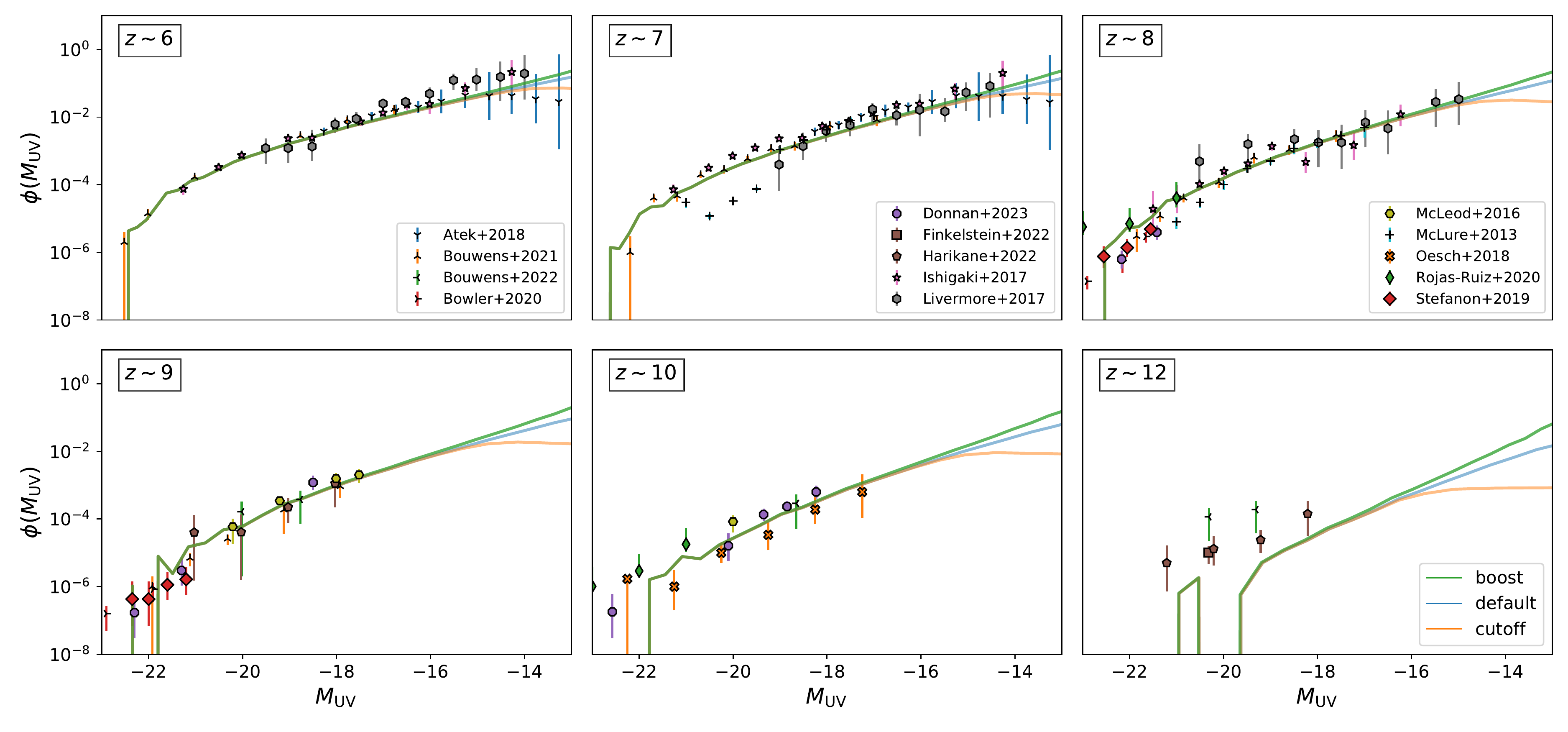}
     \caption{
     Luminosity function $\phi_{\rm UV}$ [mag$^{-1}$\,Mpc$^{-3}$], at different redshifts. The observational data points (with 1$\sigma$ error bars) have been collected from the literature \citep{McLure_+13,McLeod_16,Livermore_+17,Ishigaki_2017,Atek+18,Oesch_18,Stefanon+19,Bowler+20,Rojas-Ruiz_+20,Bouwens+21,Bouwens+22,Finkelstein_22,Donnan+23,Harikane_2023}. 
     The three solid lines correspond to the \textit{default} (blue), \textit{cutoff} (orange), and \textit{boost} (green) model presented in the text. They are characterised by different shapes of the stellar-to-halo ratio at the low-mass end.}
     \label{fig:Lumi_fct}
\end{figure*}

\section{Results}\label{sec:results}
In this section, we present new simulations of the 21-cm signal at cosmic dawn based on {\tt BEoRN} combined with a large $N$-body run. The simulations are calibrated against high-redshift observations of the ultraviolet (UV) luminosity functions, the global ionisation fraction, and the CMB optical depth measurement. We focus on three benchmark models with different astrophysical source parameters. While they all agree with the high-redshift observations considered here, they provide very different predictions regarding the 21-cm signal. 


\subsection{N-body simulation}\label{4.1}
To obtain converged 21-cm power spectra during reionisation, a large-scale simulation volume with box-size of at least $\sim 150$ cMpc is required \citep{Iliev_14_How_Large, giri2023suppressing}. Additionally, the resolution has to be sufficiently high to include haloes down to the atomic cooling limit (i.e. approximately at $M_h\sim 10^8$ $M_\odot$) as they are believed to be the first sites of star formation. Accounting for these requirements, we run a gravity-only $N$-body simulation for a 147 cMpc box with $2048^{3}$ particles, yielding a particle mass of $M_{\rm part}=1.47\times 10^{7} M_{\odot}$. We use the N-body code {\tt Pkdgrav3} \citep{Potter2017} initialising the simulation at $z=150$ based on a transfer function from {\tt CAMB} \citep{CAMB_Lewis_2000} with cosmological parameters specified in Sec.~\ref{intro}. The halo catalogue is obtained with the on-the-fly friends-of-friend halo finder embedded in {\tt Pkdgrav3} assuming a linking length of 0.2 times the inter-particle distance. We include haloes with at least 10 DM particles in our catalogue, which correspond to a minimum halo mass of $M_{\rm h, min} = 1.47\times 10^{8} M_{\odot}$. The density fields and halo catalogues are saved every 10 Myr between $z=25$ and 6. More details about the simulation and the halo catalogue can be found in Appendix~\ref{Appendix:HMF}. We generate velocity fields from the density maps following the method outlined in \citep{21cmFAST_2010}. They are used in the analysis to include RSD effects.

\begin{table}
 \begin{center}
 \renewcommand{\arraystretch}{1.3}
\centering
  \caption{
  Specific parameters for the three models. $\gamma_{3}$, $\gamma_{4}$ and $M_{\rm t}$  control the low mass tail of $f_{*}$ (Eq.~\ref{eq:f_star_functional_form}). $f_{\rm esc,0}$ and $\alpha_{\rm esc}$ determine the amplitude and power-law slope of the escape fraction of ionising photons (Eq.~\ref{eq:fesc}). $f_{\rm X}$ controls the intensity of X-ray radiation (Eq.~\ref{eq:eps_xray}). The three models share common parameters that are not displayed in the table. It includes $f_{*,0}=0.02$, $\gamma_{1}=0.49$, $\gamma_{2}=-0.61 $, $M_{\rm p}= 2.8\times 10^{11} M_{\odot}$, $c_{\rm X}=10^{40.5} $ erg~s$^{-1}$yr~$M_{\odot}^{-1}$, $\alpha_{\rm X}=1.5$, $E_{\rm min}=500 \rm eV$, $E_{\rm max}=2000  \rm eV$, $N_{\rm alpha}=9690$, $\alpha_{\rm s}=0$ , and $N_{\rm ion}=5000$ (Eqs.~\ref{eq:f_star_functional_form},\ref{eq:eps_al}, and \ref{eq:eps_xray}).}
  \setlength{\tabcolsep}{1.5\tabcolsep}
  \begin{tabular}{ccccccc} 
  \hline \hline
Model & $\gamma_{3}$ & $\gamma_{4}$ & $M_{\rm t} [M_{\odot}]$ & $f_{\rm esc,0}$ & $\alpha_{\rm esc}$ & $f_{\rm x}$\\
\hline
\textit{boost}   & 1 & 1 & 7.35$\times 10^{8}$  &  0.26 & 0  & 5  \\
\textit{default} & 0 &  0 & 0 & 0.3 & 0.2   & 1    \\
\textit{cutoff}  & 4 & -4 & 1.47$\times 10^{9}$  & 0.43  & 0.5  & 0.1 \\
  \hline \hline
  \label{tab:3Models param}
  \end{tabular}
 \end{center}
\end{table}


\subsection{Constraints from luminosity function}\label{4.2}


The UV luminosity function $\phi_{\rm UV}$ is a measure of the number density of galaxies at a given redshift, observed in a specific rest frame UV frequency band. It is determined using photometric data from telescopes, such as Spitzer \citep{Bowler+20}, the Hubble Space Telescope \citep{McLure_+13,McLeod_16,Livermore_+17,Ishigaki_2017,Atek+18,Oesch_18,Stefanon+19,Rojas-Ruiz_+20,Bouwens+21}, and, more recently, the James Webb Space Telescope \citep{Bouwens+22,Finkelstein_22,Donnan+23,Harikane_2023}. The UV luminosity function provides an independent way to constrain astrophysical source parameters that are relevant for the 21-cm signal. Note that these source properties are degenerate with the nature of dark matter \citep[e.g.][]{rudakovskyi2021constraints,dayal2023wdm}. We defer the exploration of dark matter models to the future.


Two ingredients are essential to model the UV luminosity function: the halo mass function, and a model to populate haloes with galaxies. Following \citet{Munoz_nonGaussianity_UV}, we express $\phi_{\rm UV}$ as 
\begin{equation}
  \phi_{\rm UV} = \frac{dn}{dM_{h}}\frac{dM_{h}}{dM_{\rm UV}}
  \label{eq:UV luminosity}
\end{equation}
where $dn/dM_{h}$ is the halo mass function from our simulation. The absolute UV magnitude ($M_{uv}$) is a dimensionless quantity  related to the UV luminosity $L_{\rm UV}$ [$\rm erg\,s^{-1}$] via 
\begin{equation}
  \log_{10}\left(\frac{L_{\rm UV}}{\rm erg\,s^{-1}}\right)  = 0.4 \times (51.63-M_{\rm UV}).
  \label{eq:UV absolute magnitude}
\end{equation}
Finally, the UV luminosity is related to the star formation rate via 
\begin{equation}
  \dot{M_{*}} = \kappa_{\rm UV} \times L_{\rm UV} = \frac{m_{\rm p}}{N_{\rm UV}\times E_{\rm UV}} L_{\rm UV},
  \label{eq:kappa UV}
\end{equation}
where $\kappa_{\rm UV} = 1.15\times 10^{28} \rm ~[M_{\odot}yr^{-1}\,erg^{-1}s]$ is a conversion factor from \citet{Sun_Furlanetto_2016}.

\begin{figure*}
     \centering
     \includegraphics[width =1\textwidth,clip]{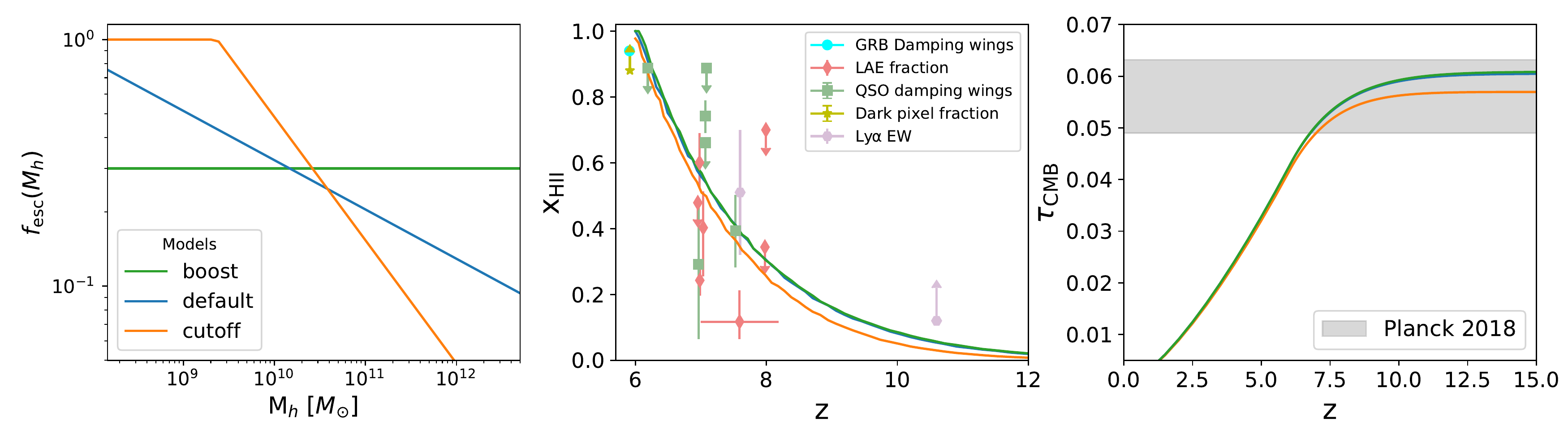}
     \caption{\textit{Left:} Escape fraction of ionising photons $f_{\rm esc}$, as a function of halo mass. We adjust the shape of $f_{\rm esc}$ to compensate for the different shapes of $f_{*}$ at small masses in order to obtain similar reionisation history in the three models. \textit{Middle:} Mean ionisation fraction history in the three models (solid lines). We gathered the observational constraints from the literature \citep{Ouchi_2010,Mortlock2011,Ono_2012,Schroeder_12,Tilvi_2014, Pentericci_2014,Totani_16,Bañados2018,Mason_2018,Hoag_2019,Ďurovčíková_20,Jung_2020,Bruton_23}. 
     \textit{Right:} CMB Thomson scattering optical depth. The shaded region corresponds to the value inferred by Planck 2018 with 1$\sigma$ uncertainty.}
     \label{fig:Results_3models_xHII}
\end{figure*}


Using our halo catalogues, we can solve Eqs.~(\ref{eq:UV luminosity}-\ref{eq:kappa UV}) to obtain the UV luminosity function at various redshifts. We tune $f_*$ to match observational data from HST, Spitzer, and JWST at different redshifts. Our best-fitting parameters are $f_{*,0}=0.05$, $\gamma_{1} = 0.49$, $\gamma_{2} = -0.61$, and $M_{\rm p} = 2.8\times10^{11} M_{\odot}$. It is worth noting that $\gamma_{1}$, $\gamma_{2}$, and $M_{\rm p}$ have the same value as in \citet{Mirocha_16}.

The low mass end of $f_{*}$ remains unconstrained due to the limited data at the faint end of $\phi_{\rm UV}$. To explore the impact of these faint galaxies on the 21-cm signal, we build three different models (\textit{boost}, \textit{default}, and \textit{cutoff}) with different shapes of $f_{*}$ at small halo masses. These models are similar to the \textit{floor}, \textit{dpl}, and \textit{steep} models from \citet{Mirocha_16}. The functional shape of their stellar-to-halo mass relation ($f_{\ast}$) are shown in Fig.~\ref{fig:3models_fstar}. The characteristics of each model are summarised below:
\begin{itemize}
   \item  \textit{Boost}: In this extreme model, $f_{*}$ starts to gradually flatten and increase again towards small halo masses below $10^{10}$ M$_{\odot}$ (see green line in Fig.~\ref{fig:3models_fstar}). We obtain this behaviour by setting $\gamma_{3}=1$, $\gamma_{4}=1$, and $M_{\rm t} = 7.35\times 10^{8} M_{\odot}$. The unusual small-scale behaviour of $f_{\ast}$ can be motivated by efficient Population III star formation in the very first galaxies.
   
   \item \textit{Default}: In this model, $f_{*}$ continues to decrease as a power law down to the smallest star-forming haloes (see blue line in Fig.~\ref{fig:3models_fstar}). The corresponding model parameters are $\gamma_{3}=0$ and $\gamma_{4}=0$. Such a model reflects a continuous suppression effect of feedback mechanisms down to the atomic cooling limit. 

   \item  \textit{Cutoff}: This model is characterised by a sharp cutoff in $f_{*}$, with no star-formation in haloes below $10^9$ M$_{\odot}$ (orange line in Fig.~\ref{fig:3models_fstar}). The functional shape is given by the parameters $\gamma_{3}=4$, $\gamma_{4}=-4$, and $M_{\rm t}=1.47\times 10^{9} M_{\odot}$. The model represents a situation where feedback effects completely shut down star formation in smaller dwarfs due to e.g. the ejection of gas outside of the potential well preventing any further star formation.
\end{itemize}
For all three models, the stellar-to-halo mass relation is truncated at $M_{\rm h}=1.47\times10^{8}M_{\odot}$ corresponding to the resolution limit of our simulation. Note that this roughly agrees with the atomic cooling limit below which gas cooling becomes inefficient (as it relies on the presence of H$_2$ molecules).

In Fig.~\ref{fig:Lumi_fct} we plot the observed UV luminosity functions (data points) together with our predictions from Eq.~(\ref{eq:UV luminosity}). The three benchmark models are shown in green (boost), blue (default), and orange (cutoff). All three provide a good fit to the data across all redshifts, differing only at magnitudes where no data is available. The noise in our curves at low magnitudes ($M_{\rm UV}<-20$), more pronounced towards the highest redshift, is explained by the small number of large haloes in our simulation (see the error bars in the HMF plot in Fig.~\ref{Appendix:HMF}). Given the simplicity of the model, the agreement between theory and observations is surprisingly good. However, a visible discrepancy starts to appear at $z=12$, where the data point ts lies above the  curves. This appearing tension might point towards a redshift dependence in the stellar-to-halo mass relation.

\begin{figure*}
     \centering
     \includegraphics[width =1\textwidth,clip]{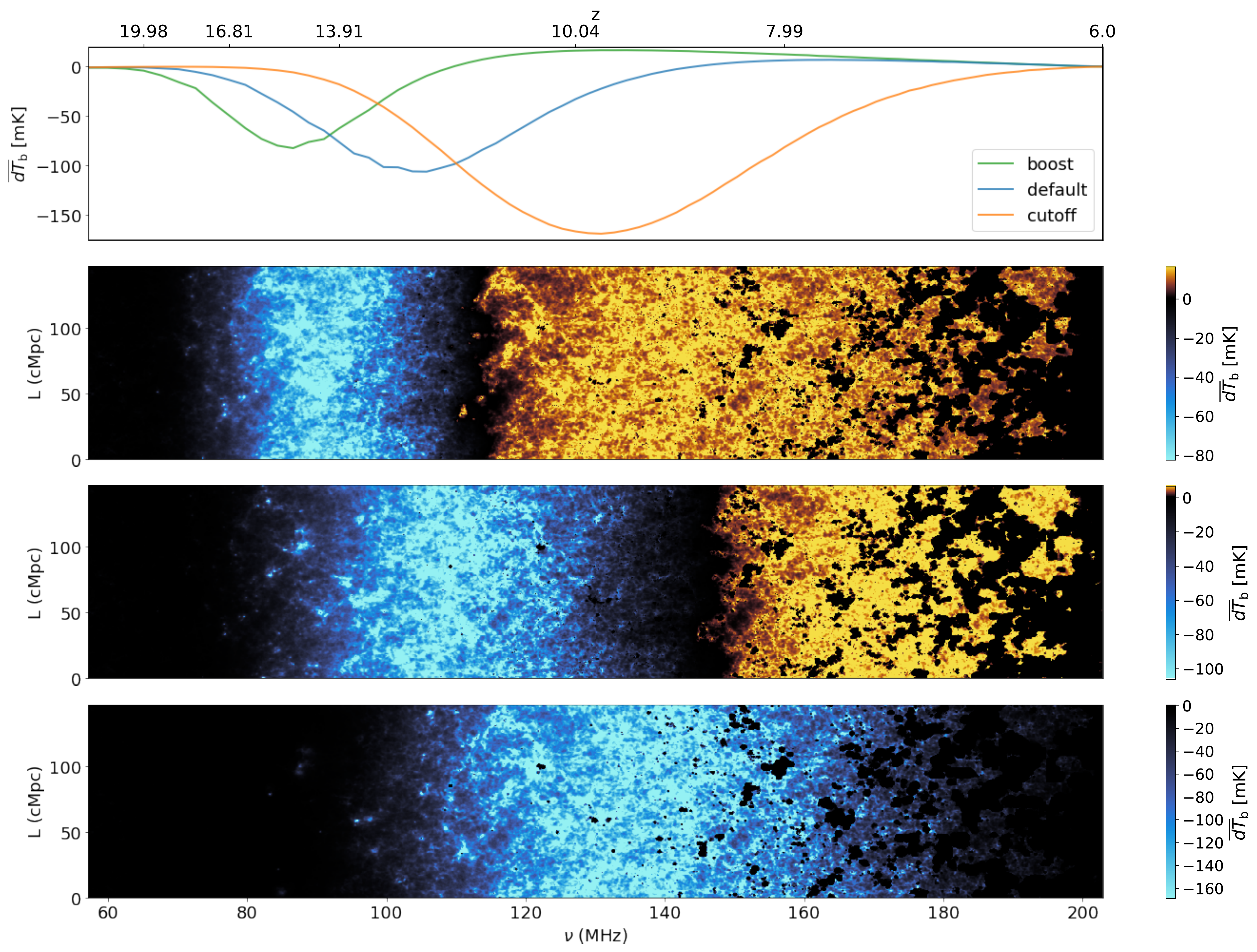}
     \caption{\textit{Upper panel:} Mean differential brightness temperature $\overline{dT}_{\rm b}$ for the \textit{boost} (green), \textit{default} (blue), and \textit{cutoff} model (orange). \textit{Lower three panels:} Lightcone images showing brightness temperature slices ($dT_{\rm b}$) as they evolve with time. RSD are included. The three images correspond to the \textit{boost}, \textit{default}, and \textit{cutoff} models (from top to bottom). The blue colour indicates regions where $dT_{\rm b}<0$ (signal in absorption). The yellow colour denotes cells of the IGM that are heated above the CMB temperature, leading to a positive value of $dT_{\rm b}$ (signal in emission). The black areas correspond to a lack of signal. The appearing patches on the right-hand side of the lightcone images indicate ionised bubbles.}
     \label{fig:Results_3models_lightcones}
\end{figure*}

\begin{figure*}
     \centering
     \includegraphics[width =1\textwidth,clip]{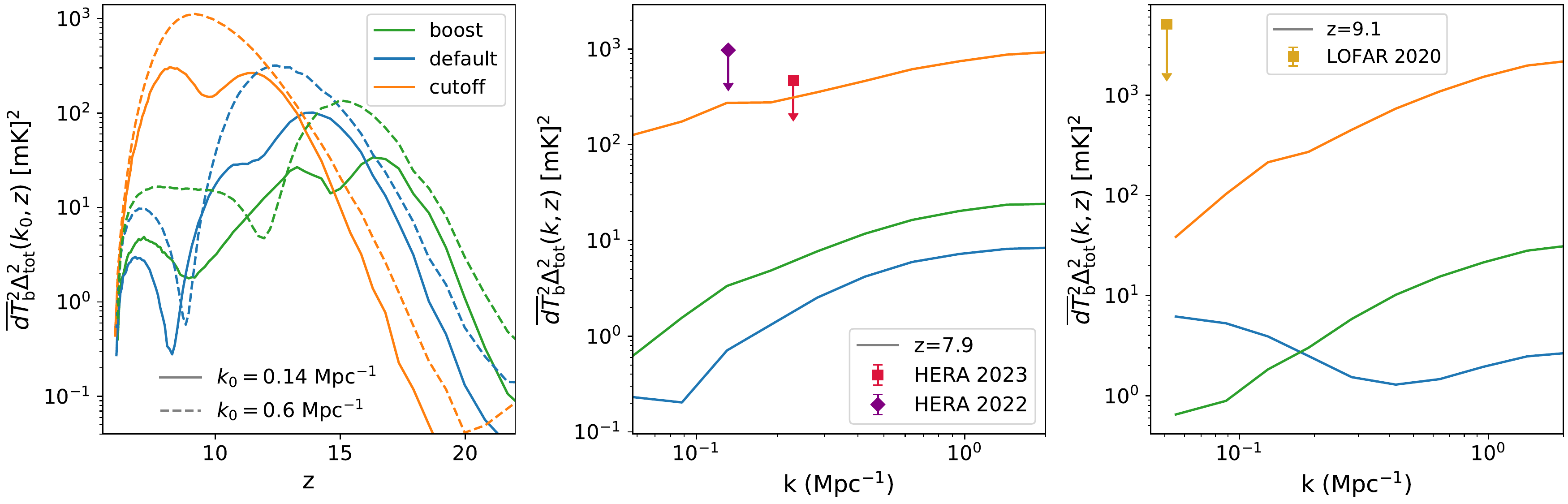}
     \caption{\textit{Left:} Dimensionless power spectrum of the $dT_{\rm b}$ field as a function of redshift, for the three models - \textit{default}, \textit{cutoff} and \textit{boost} respectively in green, blue and orange.  RSD are included. Solid and dashed lines correspond respectively to the Fourier modes $k=0.14 \rm \,Mpc^{-1}$ and $k=0.6 \rm \,Mpc^{-1}$.  \textit{Middle:} Dimensionless power spectrum at redshift $z=7.9$, as a function of scale $k$. The red square and purple diamond are the best HERA upper limits \citep{HERA_22a,HERA_22b}. 
     \textit{Right:} Dimensionless power spectrum at redshift $z=9.1$, as a function of scale $k$. The yellow square is the best LOFAR upper limit at this redshift \citep{LOFAR_z9}. }
     \label{fig:Results_3models_PS_dTb}
\end{figure*}

\subsection{Constraints from reionisation}\label{4.3}
In our framework, the timing and duration of reionisation are controlled by the product of three quantities: $f_{*}\times f_{\rm esc}\times N_{\rm ion}$. We have determined $f_{*}$ for the three models in Sec.~\ref{4.2} but the values of $f_{\rm esc}$ and $N_{\rm ion}$ remain yet to be defined. Following \citet{Park_18}, we set the number of ionising photons per stellar baryons to $N_{\rm ion}=5000$. We then vary the $f_{\rm esc}$ to guarantee the ionisation process to end at $z\sim 6$.

For the \textit{default} model, we set the parameters for the escape fraction (see Eq.~\ref{eq:fesc}) to $f_{\rm esc,0}$ = 0.3, and $\alpha_{\rm esc}=0.2$. This corresponds to a functional form in qualitative agreement with results from hydro-dynamical simulations \citep{f_esc_hydro_Taysun}. For the \textit{cutoff} model, we assume a steeper function with
$f_{\rm esc,0}=0.43$ and $\alpha_{\rm esc}=0.5$ compensating the lack of small star-forming haloes. The \textit{boost} model finally requires a small escape fraction to counteract the high star formation efficiency at small halo masses. We assume a flat function with $f_{\rm esc,0}$ = 0.26, and $\alpha_{\rm esc}=0$.

In Fig.~\ref{fig:Results_3models_xHII} we plot the escape fraction (left), the global ionization fraction (centre), and the CMB Thomson scattering optical depth (right) for all three benchmark models. The evolution of the \textit{default}, \textit{boost}, and \textit{cutoff} models are illustrated by green, blue, and orange lines, respectively.

For the global ionisation fraction shown in the middle panel of Fig.~\ref{fig:Results_3models_xHII}, we have added observational data from the literature (coloured data points). The measurements come from observations of Lyman-$\alpha$ emitters \citep{Ouchi_2010, Ono_2012, Tilvi_2014, Pentericci_2014, Mason_2018, Hoag_2019}, Lyman-$\alpha$ equivalent widths \citep{Jung_2020,Bruton_23}, dark pixel fractions \citep{McGreer_15}, as well as GRBs and QSOs damping wings \citep{Mortlock2011,Schroeder_12,Totani_16,Banados2018,Duruvcikova_20}. While they do not all agree with each other, they nevertheless provide a consistent picture of an ionisation history between $z\sim 10$ and $6$ which is well reproduced by our benchmark models. 

In the rightmost panel of Fig.~\ref{fig:Results_3models_xHII}, we indicate the Thomson optical depth measured from the Planck 2018 data \citep{Planck_2018}. 
This quantity constrains the integrated history of reionization \citep[see e.g. eq.~12 in][]{Bianco_2021}.
The predictions from our three benchmark models 
are within the $1\sigma$ contour (grey band).

\subsection{Fixing the remaining parameters}\label{4.4}
We have now fixed all the parameters connected to pre-reionisation observations. The Lyman-$\alpha$ and X-ray fluxes remain to be specified. To match the Population II stellar model from \citep{Barkana_2005}, we set the number of Lyman-$\alpha$ photons per stellar baryons to $N_{\alpha} = 9690$. We assume the spectra of Lyman-$\alpha$ radiation to be flat ($\alpha_{s} = 0$). 

For the X-ray normalisation, we follow \citep{Park_18} and normalize the flux to $c_{\rm X} = 10^{40.5}$ erg~s$^{-1}$yr~$M_{\odot}^{-1}$, in the energy range defined by $E_{\rm min} = 500 \rm \, eV$, $E_{\rm max} = 2 \rm \,  keV$, with a power-law index $\alpha_{\rm X} = 1$. By doing so, we are assuming that high-mass X-ray binaries (HMXBs) are the dominant sources of X-ray radiation during the entire heating and reionisation period and that soft X-rays with energy below 500 $\,$eV are absorbed in the interstellar medium. These parameters are consistent with the simulations from \citep{Fragos_2013,Das_2017}. However, the true normalisation and shape of the X-ray spectrum emitted by high redshift galaxies remain highly uncertain. Therefore, we select different values of $f_{\rm X}$ for our three benchmark  models. In the \textit{cutoff} model,  $f_{\rm X}$ is set to 0.1, which corresponds to a rather inefficient heating scenario \citep[similar to what is assumed in][]{Mirocha_16}. For the \textit{default} model, we assume $f_{\rm X}=1$, our default assumption for the X-ray flux. In the \textit{boost} model, $f_{\rm X}$ is set to 5 which means that the IGM is assumed to be very efficiently heated by X-ray sources.

\subsection{Resulting 21-cm signal}\label{4.5}
Given the astrophysical assumptions discussed above, we now look at the simulations of the 21-cm signal predicted by our three benchmark models. We first investigate the global signal and the corresponding light-cone maps in the following subsection. Later we study the corresponding power spectra that are expected to be observed by radio interferometers.

\subsubsection{Global signal and lightcone images}
The sky-averaged differential brightness temperature $\overline{dT_{\rm b}}(z)$ or global signal along with the lightcone maps of the three benchmark models are illustrated in  Fig.~\ref{fig:Results_3models_lightcones}.
The different stages of cosmic dawn and epoch of reionisation are clearly visible in the 21-cm signal lightcone maps shown in the bottom three rows. Regions in blue and yellow represent the signal seen in absorption and emission respectively. Black regions correspond to no 21-cm signal. The black patches at the low redshift end correspond to ionised bubbles created due to the reionisation process. The main coloured areas can be directly connected to the global signal plotted in the top panel.

The Lyman-$\alpha$ coupling epoch marks the beginning of the cosmic dawn. In the three models, the emission of Lyman-$\alpha$ photons drives the spin temperature to the kinetic temperature $T_{\rm k}$ and leads to a characteristic absorption trough. This trough is seen at $z\sim 15, 12.2$ and 9.7 with minima values of $\overline{dT_{\rm b}}\sim -82.4, -106.2$ and $-168.8$ mK, respectively in the \textit{boost}, \textit{default} and \textit{cutoff} models. These differences in the timing of the absorption trough illustrate the sensitivity of the high redshift $dT_{\rm b}$ to the tail of the SFE for small haloes with masses between $10^{8}M_{\odot}$ and $10^{10}M_{\odot}$. During this pre-heating stage, the gas temperature $T_{\rm k}$ is significantly lower than $T_{\rm CMB}$, and hence $dT_{\rm b}<0$. The transition is visible in the lightcone maps with the colour transitioning from black to blue. Note that the three models are far from the low-frequency band of EDGES observation \citep{Edges_2018}, which detected an absorption trough centred between $z\sim20$ and $z\sim15$. This hints that there is a tension between the EDGES observation and the UV luminosity function data set. As already shown by \citep{Mirocha_2019}, fitting simultaneously the UV luminosity function and the timing of the dip of the EDGES signal can only be achieved by a dramatic increase of $f_{*}$ in minihaloes (similar to our \textit{boost} model, but even more extreme). 

The next stage of our Universe observed by the 21-cm signal is known as the epoch of heating. During this stage, X-ray photons cause an increase in the $T_{\rm k}$ of the gas. If $T_{\rm k}$ rises above the CMB temperature, the signal is observed in emission ($dT_{\rm b}>0$). We see this epoch as yellow regions in the lightcone maps of Fig.~\ref{fig:Results_3models_lightcones}. Heating is triggered  efficiently in the \textit{boost} and \textit{default} models, due to both the values of $f_{\rm X}$ and the high SFE at small halo masses. In the \textit{cutoff} model, the mean $T_{\rm k}$ never goes above $T_{\rm cmb}$. Such a signal corresponds to a ``cold reionisation'' scenario, in which ionised bubbles grow surrounded by a cold neutral IGM. Our results for the \textit{cutoff} model are consistent with those from \citep{Mirocha_16}, who assumed similarly inefficient heating parameters and obtained the same type of global signal.


During the epoch of reionisation, UV photons start to ionise the intergalactic medium (IGM) forming bubbles around sources. These ionised bubbles appear as black patches in the lightcone images and start to grow at around $z\sim10$, percolating around $z\sim 8$ until they fill the entire box volume by $z\sim 6$. The morphology of ionisation depends on the halo distribution, the star-formation efficiency, the UV photon production, and the escape fraction. Note that the three models show similar morphologies of their ionisation maps. This can be explained by the fact that we have adjusted the shape of the escape fraction to compensate for the differences in star-formation efficiency.

\subsubsection{Power Spectrum}
In the leftmost panel of Fig.~\ref{fig:Results_3models_PS_dTb}, we show the redshift evolution of the dimensionless power spectra of the $dT_{\rm b}$ fields for the three benchmark models following the same colour scheme as before. Solid and dashed lines correspond to the different wave modes $k=0.14 \rm Mpc^{-1}$ and $k=0.6 \rm Mpc^{-1}$. As expected, the three spectra display distinct peaks and troughs, which correspond to the epochs when the $dT_{\rm b}$ field is dominated by fluctuations of the Lyman-$\alpha$ coupling ($x_{\alpha}$), the temperature ($T_{\rm k}$), and the ionisation ($x_{\rm HII}$) field.

The Lyman-$\alpha$ peak occurs roughly at the midpoint of Lyman-$\alpha$ coupling, when $dT_{\rm b}$ fluctuates between regions still coupled to the CMB ($dT_{\rm b}=0$), and regions in absorption coupled to the kinetic temperature ($dT_{\rm b}<0$). It is located at $z\sim16.5,14,11.5$ in the \textit{boost}, \textit{default}, and \textit{cutoff} models, respectively. The Lyman-$\alpha$ peak is followed by a gap, which occurs when the mean $dT_{\rm b}$ is minimal (respectively at $z\sim14.6,11.6,9.7$). This is due to the negative cross-correlation between the $x_{\rm \alpha}$ and $T_{\rm k}$ fields. 

After that point, the three models can be divided into two categories: (i) In the \textit{boost} and \textit{default} models, the spectra have a heating peak. It occurs roughly when the mean $dT_{\rm b}$ is halfway from reaching its maximum positive value. The peak is followed by a second trough when the mean $dT_{\rm b}$ is maximal. The spectra then hit a third peak, when respectively 50$\%$ and $60\%$ of the universe is ionised.
(ii) In the \textit{cutoff} model, the spectrum has no heating peak. Heating is inefficient and proceeds very slowly: 
$dT_{\rm b}$ fluctuates between cold regions with negative values and growing ionised regions with no signal ($dT_{\rm b}=0$). The power spectrum hits a single peak when $30\%$ of the universe is ionised. It reaches very high values ($\sim 3\times10^{2} \rm mK^{2}$ at $k=0.14 \rm \, Mpc^{-1}$), which reflects the strong contrast between neutral regions with $dT_{\rm b}\sim-150~\rm mK$ and ionised region with $dT_{\rm b}=0 \,\rm mK$.

 

In the middle and rightmost panels of Fig.~\ref{fig:Results_3models_PS_dTb}, we plot the power spectra as a function of $k$-modes at the distinct redshift values of $z=7.9$ and $z=9.1$. Next to our results, we plot the current upper limits on the 21-cm power spectrum HERA \citep[][which are shown as red square and purple diamond markers respectively]{HERA_22a,HERA_22b}, 
and LOFAR observations \citep[][which is shown as a yellow square]{LOFAR_z9}. 
While the upper limits are still significantly higher than the bulk of predictions, the \textit{cutoff} model comes close to the limits from HERA (intersecting with the $1\sigma$ error bar). This finding is in agreement with previous work  \citep{HERA_interpretation_Abdurashidova_2022,LOFAR_constraints_Ghara_2020,MWA_constraints_Ghara_21, MWA_constr_Greig} in which cold reionisation scenarios were shown to be disfavoured.


\section{Conclusions}\label{sec:conclusion}



In the near future, radio interferometers such as the Square Kilometre Array (SKA) telescope are expected to provide first observations of the 21-cm signal from the epoch of reionisation and cosmic dawn. These observations will yield new insights into the properties of the very first galaxies and they will help stress-testing the $\Lambda$CDM model in a currently un-probed regime.

In order to interpret and understand the 21-cm signal at cosmic dawn, fast and accurate prediction methods are required. In this paper we introduce the \textit{Bubbles during the Epoch of Reionisation Numerical simulator} ({\tt BEoRN}), a new simulation framework to produce maps and lightcone images of the 21-cm signal. {\tt BEoRN} is designed as a user-friendly and modular, open-source code written entirely in {\tt Python}. It can produce 3-dimensional maps of the 21-cm brightness temperature $dT_{\rm b}$ from cosmic dawn to reionisation, including Lyman-$\alpha$ coupling, heating, and ionisation of neutral hydrogen by galaxies.

\beorn is based on the concept of overlying flux profiles around sources and follows the basic methodology developed in \citet{HaloModel_Paper,HaloModel_reio}. The method works by pre-calculating individual profiles of Lyman-$\alpha$, X-ray, and ionising photon fluxes for different halo masses, assuming a given stellar-to-halo mass relation and including photon redshifting as well as look-back effects due to the finite speed of light. The resulting profiles are painted on a 3-dimensional grid using halo positions and density fields from a $N$-body simulation. Overlaps of ionising bubbles are dealt with by re-distributing excess photons at the bubble boundaries. 


We validate {\tt BEoRN} by comparing it to the semi-numerical algorithm {\tt 21cmFAST}. The two codes agree reasonably well for all relevant redshifts and $k$-modes (as shown in Figs.~\ref{fig:21cmFAST_BEoRN_Comparison} and \ref{fig:lightcones_FAST_BEORN}). However, during the epoch of bubble growth, \beorn predicts a larger number of smaller, more irregular ionised patches compared to {\tt 21cmFAST}. This results in a significant difference in amplitude of the power spectrum at large scales, in qualitative agreement with earlier comparisons between semi-numerical codes and radiative transfer simulations \citep{Zahm_11_Comparison_RT_SemNum,Majumdar_14_Comparison_RT_SemNum}. 


After validating the code, we run a large $N$-body simulation (with a box length of 147 cMpc and $N=2048^3$ particles) using the resulting density grids and the halo catalogues as input fields for \beorn. We study three benchmark models with vastly different choices for the astrophysical parameters, that are, however, all selected to reproduce recent observations of the UV luminosity functions, the global ionisation fraction, and the Thomson optical depth measurement. The three benchmark models are characterised by different stellar-to-halo relations ($f_{*}$) at small halo masses, different escape fractions for the ionising photons, and different normalisation factors for the emitted X-ray flux. They lead to strongly different 21-cm signals ranging from a cold reionisation scenario with a deep absorption trough at late times to an emission-dominated scenario with only a shallow absorption trough at high redshift. The global signal evolution and lightcone images of each benchmark model are shown in Fig.~\ref{fig:Results_3models_lightcones}. 

In this work, we have focused on the global effects of feedback, specifically Lyman-Werner and radiative feedback, on star formation within dark matter haloes. However, it is important to note that this feedback process is expected to vary with position and time \citep[e.g.,][]{dixon2016large, Hutter_2021}. We have incorporated these calculations into \beorn, and a comprehensive analysis will be presented in a forthcoming publication.

The results from our three benchmark runs confirm that current observations from the period of reionisation and cosmic dawn allow for a large variety of different scenarios that can be studied with the 21-cm global and clustering signal. This shows that upcoming radio experiments will provide crucial support for our understanding of star and galaxy formation during the first billion years after the Big Bang.


\section*{Acknowledgements}
We thank Douglas Potter and Joachim Stadel for 
helping with the {\tt Pkdgrav3} code. We thank Raghunath Ghara and Garrelt Mellema for useful discussions. TS would like to express his gratitude to Deniz Soyuer for his proofreading, and for the daily inspiration he provides. This study is supported by the Swiss National Science Foundation via the grant {\tt PCEFP2\_181157}.
Nordita is supported in part by NordForsk.

\section*{Data Availability}
The data underlying this article will be shared on reasonable request to the corresponding author.





\bibliographystyle{mnras}
\bibliography{example} 



\newpage

\appendix

\section{Speeding up the bubble overlap}\label{Appendix:overlapp}

\begin{figure}
     \centering
     \includegraphics[width =0.48\textwidth,clip]{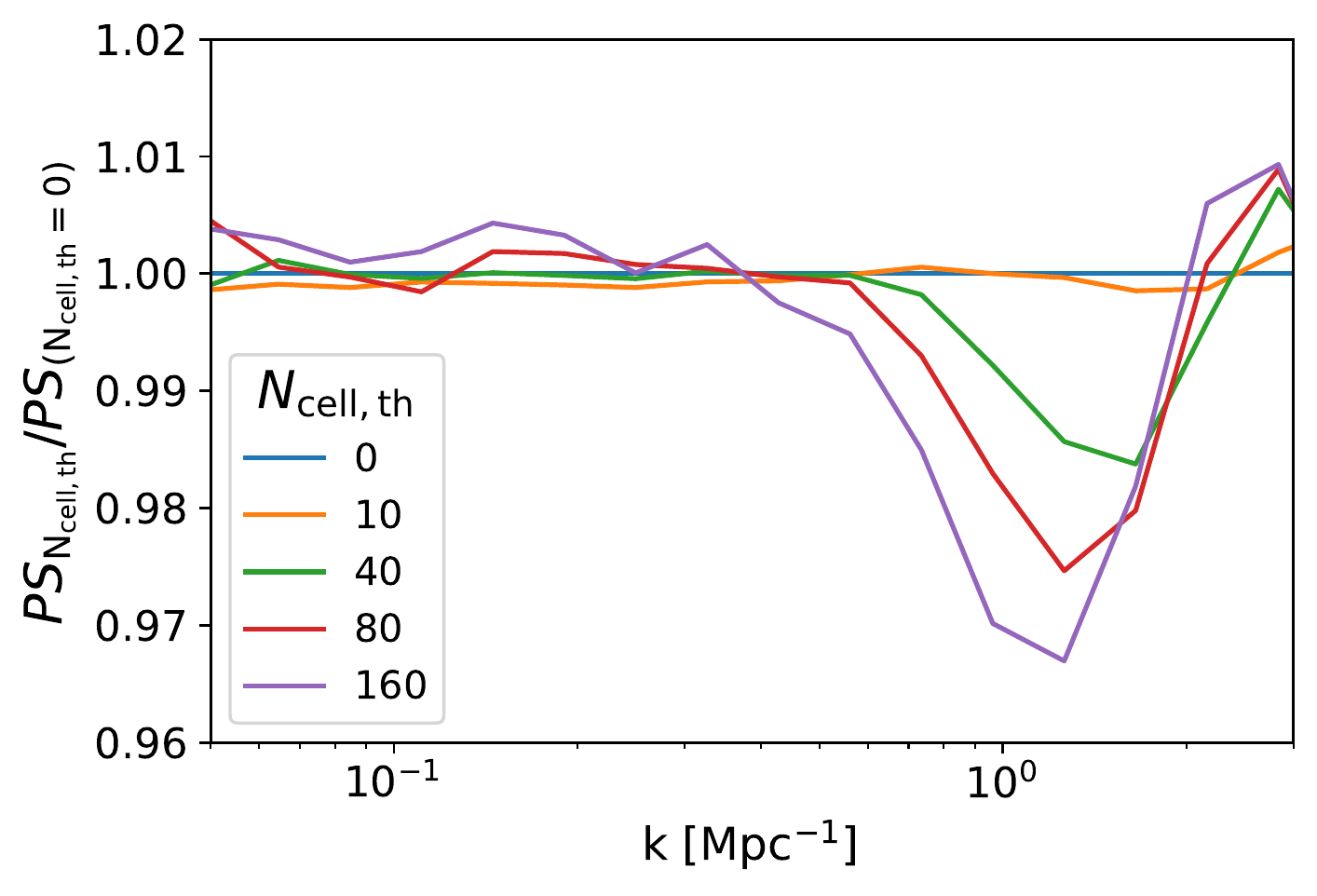}
     \caption{Effect of varying the threshold parameter $N_{\rm cell, th}$ on the power spectrum of the $x_{\rm HII}$ field. $N_{\rm cell, th}$ is used to speed up the bubble overlap procedure (see Appendix.~\ref{Appendix:overlapp}). The y-axis shows the ratio between the PS of the $x_{\rm HII}$ field obtained for different values of $N_{\rm cell, th}$, with respect to the PS of the fiducial $x_{\rm HII}$ field, obtained for $N_{\rm cell, th} = 0$. Up to $N_{\rm cell, th}=40$ (green curve), the error induced on the power spectrum is less than 1\% at the relevant scales ($0.1~{\rm Mpc^{-1}}<k<1~{\rm Mpc^{-1}}$). }
     \label{fig:Varying_Ncell}
\end{figure}


In the main text, we mentioned that redistributing overionised cells that occur in regions where bubbles overlap can be computationally expensive. The number of ionised "islands" or connected ionised regions varies based on several factors such as the number of haloes, the astrophysical model, the grid resolution, and the mean ionisation fraction. However, the majority of these islands are very small overionised regions, typically consisting of 10 pixels or less, even though they contribute very little to the global ionisation fraction (approximately 1\%). To speed up the bubble overlap step while sacrificing some small-scale precision, we introduce a free parameter called $N_{\rm cell, th}$, which represents the "threshold cell number." This parameter enables the user to simultaneously treat all the ionised islands with fewer than $N_{\rm cell, th}$ pixels when spreading the excess ionisation fraction they contain.

To illustrate this process, consider a $x_{\rm HII}$ field on a grid, with 100 separate and unconnected ionised islands that contain overlaps or overionised cells. If $N_{\rm cell, th}=0$, the algorithm distributes the excess ionisation fraction to neighbouring cells for each island individually. However, if $N_{\rm cell, th}=10$, all islands with less than 10 pixels are treated together. This means that the sum of the excess ionisation fraction in these islands is spread evenly over their combined borders.

To speed up the code, one can increase the value of $N_{\rm cell, th}$, but at the cost of precision. However, for values of $N_{\rm cell, th}$ up to a certain threshold, the impact on the 21-cm signal at the relevant scales ($0.1<k<1$ $\rm Mpc^{-1}$) remains negligible. To evaluate this effect, we generate an $x_{\rm HII}$ field on a $256^{3}$ grid with a length of 147 cMpc and a halo catalogue containing 7.5 million halos, at redshift $z=7$, from our {\tt Pkdgrav3} simulation. The field has a mean ionisation fraction $\bar{x}_{\rm HII}=0.5$. We repeat the bubble overlap procedure with varying values of $N_{\rm cell, th}$ and compare the power spectra of the $x_{\rm HII}$ fields after redistributing the overionised cells. The results are shown in Fig.~\ref{fig:Varying_Ncell}, where we plot the ratio between the $x_{\rm HII}$ power spectra for different values of $N_{\rm cell, th}$ and the fiducial value ($N_{\rm cell, th}=0$).

After painting the ionised bubbles, we obtain 43,132 ionised isolated islands (or connected regions). The procedure takes 52 minutes in total to go through each overionised region individually and spread the excess ionisation fraction. However, we can group together all the islands with less than 10, 40, 80, or 160 pixels to speed up the process. This reduces the time required to respectively 6, 2, 1.42, and 1.18 minutes. As shown in Fig.\ref{fig:Varying_Ncell}, the error on the power spectrum remains negligible up to $N_{\rm cell, th} = 40$ (less than $1\%$), and even for $N_{\rm cell, th} = 160$, it is less than 3\% between $0.1<k<1$ $\rm Mpc^{-1}$.

In summary, the parameter $N_{\rm cell, th}$ can be adjusted according to the required precision. Note that it scales with the total number of pixels on the grid.

\section{Convergence test}\label{Appendix:Convergence}

\begin{figure*}
     \centering
     \includegraphics[width =1\textwidth,clip]{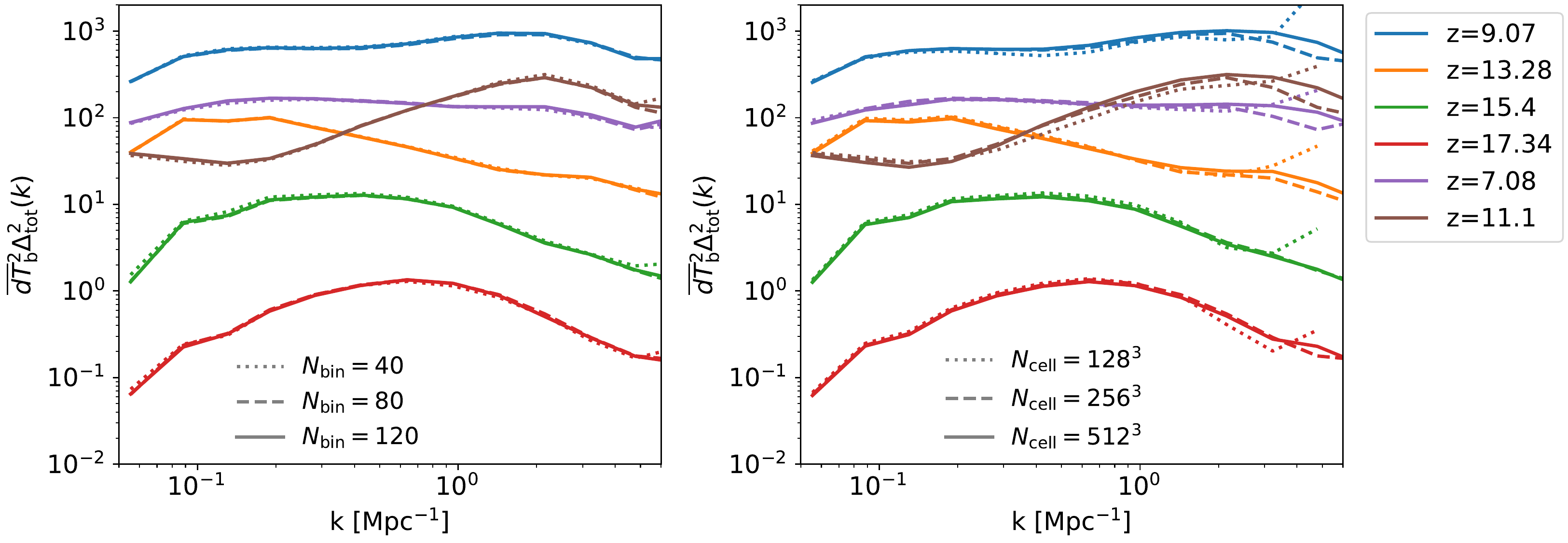}
     \caption{Halo mass and spatial resolution convergence check for {\tt BEoRN}. The underlying astrophysical model is similar to the \textit{cutoff} model from Sec.~\ref{sec:results}. \textit{Left panel:} 21-cm dimensionless power spectra at various redshifts, varying the total number of halo mass bins $N_{\rm bin}$. \textit{Right panel:} Same as the left panel, but varying the total number of grid pixels $N_{\rm cell}$. }
     \label{fig:Convergence test}
\end{figure*}

We carry out a convergence check by varying the number of halo mass bins $N_{\rm bin}$ used to compute the profiles and the number of grid pixels $N_{\rm cell}$ when running {\tt BEoRN}. We perform this comparison for a model similar to the \textit{cutoff} model of Section~\ref{sec:results} (the exact parametrization is not relevant for the purpose of this convergence test). Initially, we vary the number of mass bins $N_{\rm bin}$. Our halo mass range spans 6.5 orders of magnitude in mass. We run {\tt BEoRN} with 40, 80 and 120 bins in total on a grid with $256^{3}$ pixels and present the resulting power spectra as dotted, dashed and solid lines in the left panel of Fig.~\ref{fig:Convergence test}. $N_{\rm bin}= 80$ provides well-converged curves over the scales of interest at every redshift. Since the mass range depends on the resolution of the halo catalogues and is left free to the user, the relevant quantity to specify here is the number of mass bins per order of magnitude in halo mass. 80 bins spanning 6.5 orders of magnitude in mass is roughly equivalent to 12 bins per dex. Consequently, the user can select the minimum and maximum halo mass according to its halo catalogue and establish the number of mass bins required to achieve converged results.

Subsequently, we vary the number of pixels $N_{\rm cell}$ and present our results in the right panel of Fig.~\ref{fig:Convergence test} for $N_{\rm cell}= 128^{3}, 256^{3}, 512^{3}$ (corresponding to dotted, dashed and solid lines). The computing time scales with the number of pixels due to the FFT convolution to put profiles on the grid. $N_{\rm cell}= 256^{3}$ yields relatively well-converged results. However, if a sub-percent precision is necessary, a higher number of mass bins and finer grid resolution may be preferable.

\section{The mean free path of ionising photon in excursion-set methods}\label{Appendix:mfp}

To model the presence of unresolved dense small-scale absorbers inside ionised regions, semi-numerical excursion-set (ES) methods have introduced the parameter $R_{\max}$ \citep{Furlanetto_Oh_2005,Zahm_11_Comparison_RT_SemNum,Alvarez_2012}. $R_{\max}$ is meant to characterise the mean free path (MFP) of ionising photons inside ionised regions. It sets the value of the maximum scale over which the density and ionising photon fields are smoothed during the excursion-set procedure. In other words, a photon emitted by a source of ionising radiation cannot contribute to ionising a point that is at a distance larger than $R_{\max}$ from the source centre.



In BEoRN, the mean free path of ionising photons is not a parameter. Instead, distances travelled by ionising photons are smoothly distributed. This means that in regions with clustered sources surrounded by large ionised bubbles, ionising photons can naturally travel larger distances (via the redistribution of overionised cells) than around isolated sources surrounded by neutral gas. The only parameter in BEoRN that could have a similar effect to $R_{\max}$ is the clumping factor $C$ (see Eq.~\ref{eq:Stromgren Sphere}). However, it is unclear if there exists a mapping between $R_{\max}$ and $C$, or even if the effect of $R_{\max}$ can be reproduced by varying $C$. Therefore, there is a fundamental difference in modelling between BEoRN and ES methods that rely on $R_{\max}$.

To improve the accuracy of modelling the gradual absorption of ionising photons within ionised regions, \cite{Davies&Furlanetto} proposed a new method that accounts for the smooth exponential attenuation of the ionising flux, resulting in a scale-dependent ionisation threshold or filter. This method differs from traditional ES methods that use a sharp maximum scale ($R_{\max}$) for smoothing the density and ionising photon fields. The power spectrum predictions of the two methods differ, as we observed in our comparison of {\tt 21cmFAST} and {\tt BEoRN}.

\begin{figure}
     \centering
     \includegraphics[width =0.4\textwidth,trim=0cm 0.cm 0cm 0.0cm,clip]{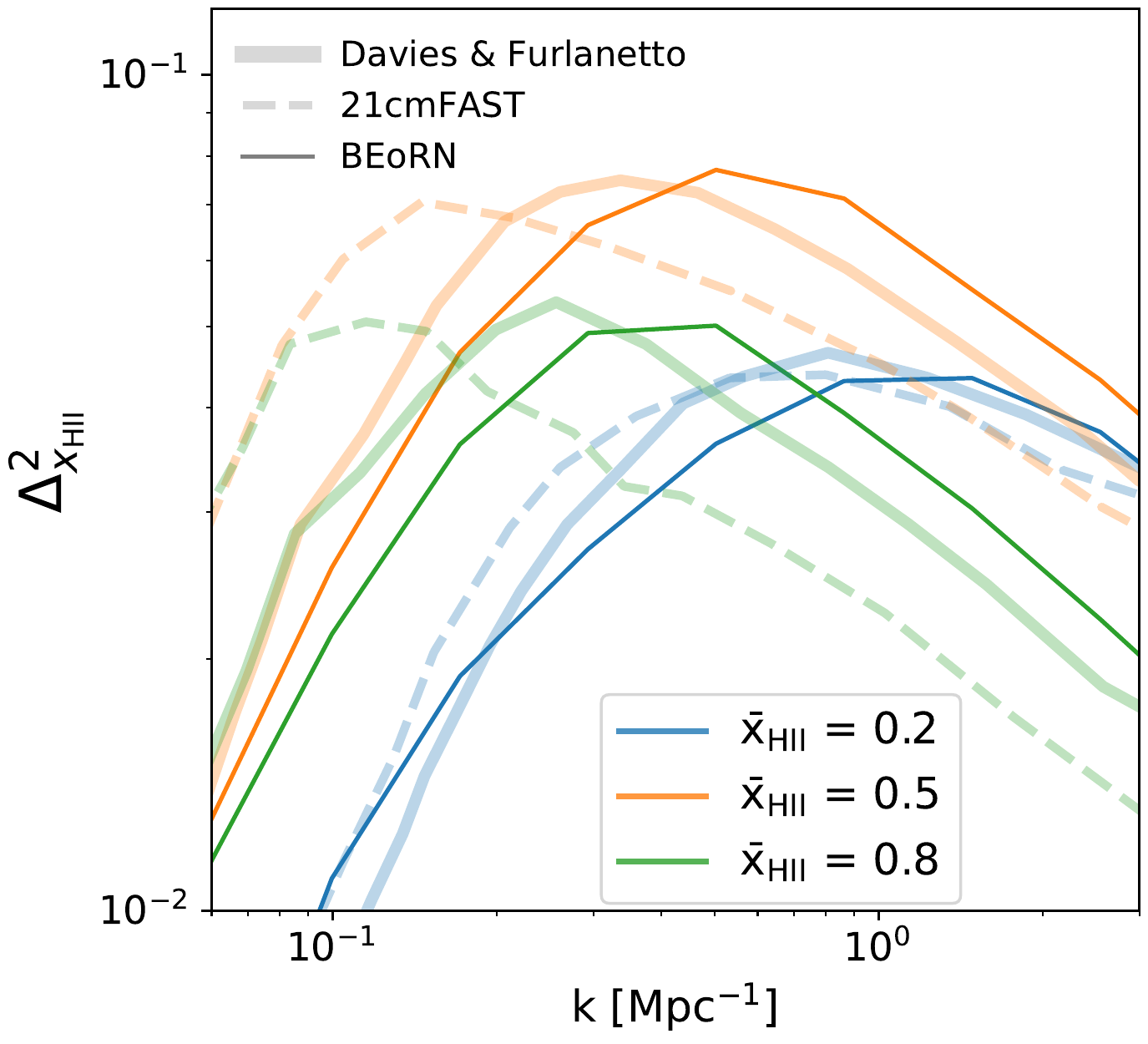}
     \caption{\textit{Upper panel:} dimensionless power spectrum $\Delta_{\rm XHII}$ of the ionisation fraction field, as a function of Fourier mode $k$ [$\rm Mpc^{-1}$]. We compare the prediction from {\tt BEoRN} (thin solid lines) with \citet{Davies&Furlanetto} (solid thick lines) and the standard approach in {\tt 21cmFAST} (dashed lines). The latter data is taken from fig.~7 of \citet{Davies&Furlanetto}. The three colours (blue, orange and green) correspond to different redshifts, $z=7.5,7,6.5$, with mean ionisation fractions, $x_{\rm HII}=0.2,0.5,0.8$ respectively.
     }
     \label{fig:Comparison_Davies}
\end{figure}

To compare {\tt BEoRN} to \cite{Davies&Furlanetto}, we use their MFP-$\epsilon(r)$ approach, which is relevant when sources of ionising photons are spatially resolved, as in {\tt BEoRN}. We run {\tt BEoRN} on top of our {\tt Pkdgrav3} halo catalogue using the same model as \cite{Davies&Furlanetto} with a constant ionising efficiency and a minimum star-forming halo of $10^{9} M_{\odot}$. We tune the number of ionising photons to achieve the same ionisation fractions as \cite{Davies&Furlanetto} at redshifts $z=7.5, 7, 6.5$ with mean ionisation fraction $x_{\rm HII}=0.2, 0.5, 0.8$. We then compare the resulting dimensionless power spectra of the $x_{\rm HII}$ fields $\Delta^2_{x_\mathrm{HII}}$ and plot them in Fig.~\ref{fig:Comparison_Davies}. The thick and dashed lines in the figure are taken from \cite{Davies&Furlanetto}, while the thin line is from {\tt BEoRN}.

The smooth MFP implementation leads to a shift of power from large to small scales, which is in good agreement with {\tt BEoRN} and is a significant improvement over traditional ES predictions. The match between {\tt BEoRN} and \cite{Davies&Furlanetto} is better, despite relying on different simulations and halo catalogues, with a maximum deviation of approximately 30\%.



\section{Halo mass function from our simulation}\label{Appendix:HMF}

\begin{figure}
    \centering
    \includegraphics[width =0.4\textwidth,trim=0cm 0.cm 0cm 0.0cm,clip]{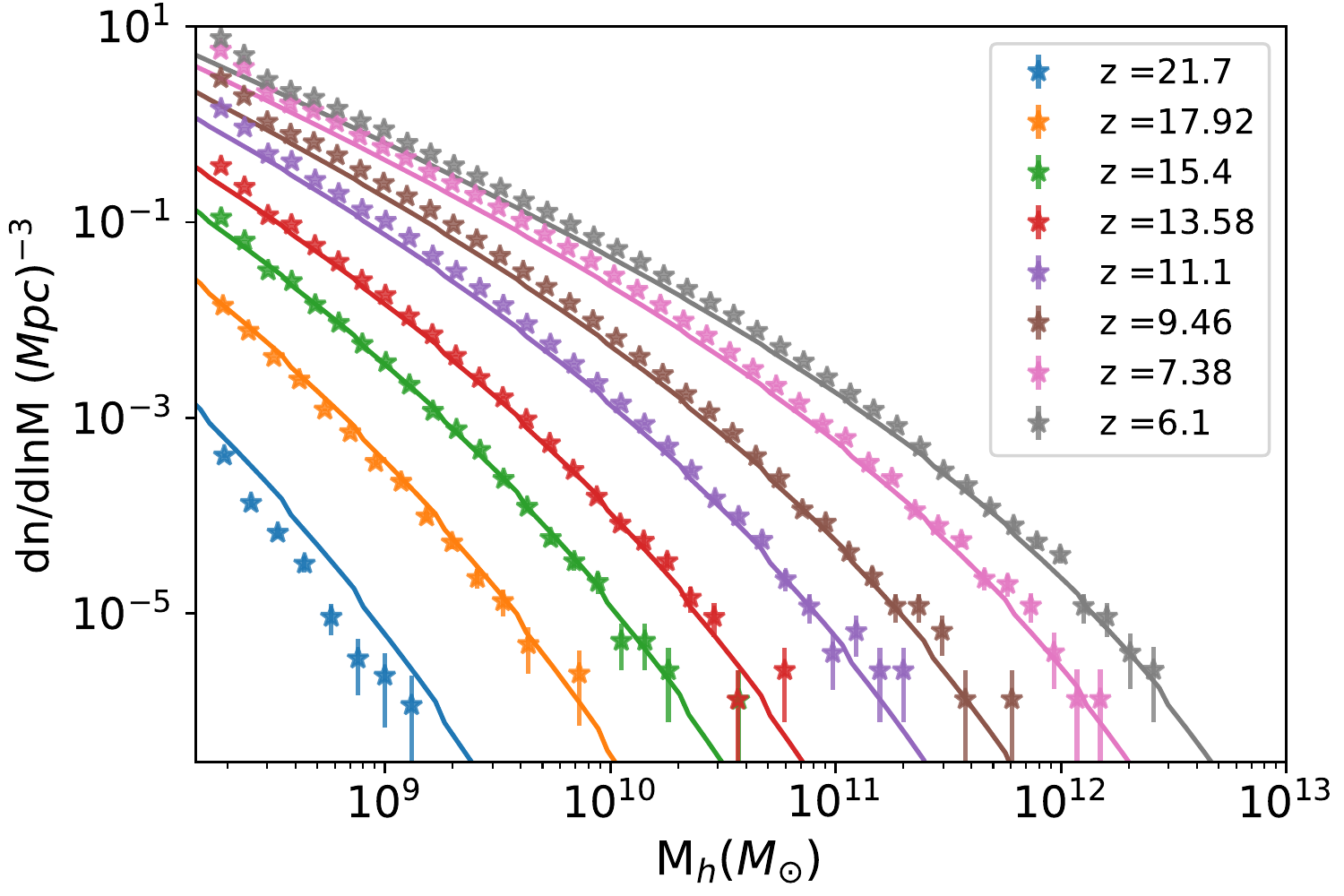}
    \caption{Halo mass function from our {\tt Pkdgrav3} N-body run (see section~\ref{4.1} for details on the simulation) with 2048$^{3}$ particles, and box size 147cMpc, at various redshifts. Solid lines correspond to a Sheth-Tormen mass function using a sharp-k filter (see Appendix.~\ref{Appendix:HMF} for the parameter values). Error bars correspond to Poisson errors in each bin.}
    \label{fig:HMF}
\end{figure}

The halo catalogues in our simulations, which consist of 2048$^{3}$ particles, were produced using the on-the-fly friend-of-friend algorithm in {\tt Pkdgrav3}. We used a linking length of 0.2 times the initial mean interparticle distance and included only haloes with at least 10 DM particles. To ensure transparency, we present the binned Halo Mass Function (HMF) from our catalogue, along with Poisson error bars, at various redshifts between $z\sim21$ and $z\sim6$, in Fig.~\ref{fig:HMF}. We also show our best-fit analytical mass function obtained through the extended Press-Schechter formalism \citep{PS, Bond_Estathiou_Kaiser_Exc_set, STellipsoidalcollapse1999}. Our Sheth-Tormen mass function with a sharp-k filter provides a good fit to the HMF over the entire range of redshifts. The solid lines in Fig.~\ref{fig:HMF} correspond to the Sheth-Tormen mass function using the following sharp-k HMF parameters: $c=2.7$, $p=0.3$, $q=1$, $\delta_{\rm c}=1.675$, and $A=0.322$ (see \citealt{Schaeffer_DAO} for parameter definitions). Note that the analytical curve slightly underestimates the halo abundance at the lowest redshifts and overestimates it at the highest redshift. Tuning the value of the parameter $c$ could improve the fit at a single redshift.

\label{lastpage}
\end{document}